 \documentclass[twocolumn]{aastex631}
\shorttitle{A Light Curve Model of V339 Del}
\shortauthors{Hachisu, Kato, \& Matsumoto}

\begin{document}

\title{A multiwavelength light curve model of the classical nova V339 Del: \\
A mechanism for the coexistence of dust dip and supersoft X-rays}


\author[0000-0002-0884-7404]{Izumi Hachisu}
\affil{Department of Earth Science and Astronomy, 
College of Arts and Sciences, The University of Tokyo,
3-8-1 Komaba, Meguro-ku, Tokyo 153-8902, Japan} 
%%%\email{hachisu@ea.c.u-tokyo.ac.jp}
\email{izumi.hachisu@outlook.jp}

%%%%\and

\author[0000-0002-8522-8033]{Mariko Kato}
\affil{Department of Astronomy, Keio University, 
Hiyoshi, Kouhoku-ku, Yokohama 223-8521, Japan} 
%%\email{mariko@educ.cc.keio.ac.jp}

%%%\and

\author[0000-0002-5277-568X]{Katsura Matsumoto} \affil{Astronomical Institute,
Osaka Kyoiku University, Kashiwara-shi, Osaka 582-8582, Japan}
%%\email{matsumoto@cc.osaka-kyoiku.ac.jp}

%\and
%
%\author{Rosario Gonz\'alez-Riestra} \affil{XMM Science Operation
%Centre, ESAC, P.O. Box 78, 28091 Vallanueva de la Ca\~nada, Madrid,
%Spain}
%\email{Rosario.Gonzalez@sciops.esa.int}

%\author{Angelo Cassatella}
%\affil{INAF, Istituto di Fisica dello Spazio Interplanetario,
%Via del Fosso del Cavaliere 100, 00133 Rome, Italy}
%\email{cassatella@fis.uniroma3.it}

%\and
%
%\author{Taichi Kato}
%\affil{Department of Astronomy, Kyoto University, 
%Sakyo-ku, Kyoto 606-8502, Japan} 
%\email{tkato@kusastro.kyoto-u.ac.jp}

%\altaffiltext{1}{Visiting Astronomer, Cerro Tololo Inter-American Observatory.
%CTIO is operated by AURA, Inc.\ under contract to the National Science
%\altaffiltext{2}{Society of Fellows, Harvard University.}
%\altaffiltext{3}{present address: Center for Astrophysics,
%    60 Garden Street, Cambridge, MA 02138}
%\altaffiltext{4}{Visiting Programmer, Space Telescope Science Institute}
%\altaffiltext{5}{Patron, Alonso's Bar and Grill}


\begin{abstract}
The classical nova V339 Del 2013 is characterized by a 1.5 mag dip of the
$V$ light curve owing to a dust shell formation, during which soft
X-ray emissions coexist.  We present Str\"omgren $y$ band
light curve, which represents continuum emission, not influenced
by strong [\ion{O}{3}] emission lines.
The $y$ light curve monotonically decreases in marked contrast to 
the $V$ light curve that shows a 1.5 mag dip.  
We propose a multiwavelength light curve model that
reproduces the $y$ and $V$ light curves as well as the gamma-ray
and X-ray light curves.  In our model, a strong shock arises far outside
the photosphere after optical maximum,
because later ejected matter collides with earlier ejected gas.
Our shocked shell model explains optical
emission lines, H$\alpha$, hard X-ray, and gamma-ray fluxes.  
A dust shell forms behind the shock that suppresses [\ion{O}{3}].
This low flux of [\ion{O}{3}] shapes a 1.5 mag drop in the $V$ light curve.
Then, the $V$ flux recovers by increasing
contribution from [\ion{O}{3}] lines, while the $y$ flux does not.
However, the optical depth of the dust shell is too small
to absorb the photospheric (X-ray) emission of the white dwarf.
This is the reason that a dust shell and a soft X-ray radiation coexist. 
We determined the white dwarf mass to be $M_{\rm WD}=1.25\pm 0.05~M_\sun$ and
the distance modulus in the $V$ band
to be $(m-M)_V=12.2 \pm 0.2$;
the distance is $d= 2.1\pm 0.2$ kpc for the reddening of $E(B-V)=0.18$.
\end{abstract}




\section{Introduction}
\label{introduction}
A classical nova is an explosion of a hydrogen-rich envelope on a
mass-accreting white dwarf (WD) in a binary.  Hydrogen ignites to
trigger an outburst when the mass of the envelope reaches a critical value
\citep[e.g.,][for a recent fully self-consistent nova explosion 
model]{kat22sha}.

\citet{hac22k} theoretically found a strong shock formation far outside
the photosphere in nova outbursts based on \citet{kat22sha}'s nova explosion
model.  A strong shock inevitably arises after the optical 
maximum because the photospheric wind velocity increases after the
optical maximum so that the wind ejected later catches up with
the wind ejected earlier.  Thus, a strong shock naturally arises
far outside the photosphere after the optical maximum.  

Nova ejecta with a shock show a rich variety of emission/absorption line
systems.  Expanding ejecta with different velocities makes multiple
velocity emission/absorption line systems \citep{mcl42}.
Also high energy emissions, such as hard X-rays and gamma-rays are
expected from the shock \citep[e.g.,][]{fri87, muk01i, mar18dj}. 
\citet{hac23k} explained how and where McLaughlin's optical multiple
emission/absorption line systems arise in classical novae. 
%as well as hard and supersoft X-rays, and gamma-rays, 

In the present work, we elucidate the nature of the classical nova 
V339 Del based on our nova shock model \citep{hac22k, hac23k}.
The outburst of V339 Del was optically discovered by K. Itagaki at 6.8 mag
on UT 2013 August 14.584 \citep[$=$JD 2,456,519.084,][]{nak13s}.
% The new object is located at R.A. = 20h23m30s.73, 
% Decl. = +20o46'04''.1 (equinox 2000.0).
% 2013 August 1 = 2456505.5
% 16.45+/-0.06  --> JD 2456520.95-0.06+0.06 --> 520.89,521.01 
% 18.4+/-0.11   --> JD 2456522.90-0.11+0.11 --> 522.79,523.01
%
Immediately after the discovery, it was well observed in multiwavelength
bands, especially in photometry, because of its brightness 
(a naked-eye nova; $V\sim 4.4$ at peak). 
See Section \ref{observation_summary} for an observational summary.

V339 Del is also characterized by 
(1) the third GeV gamma-ray detected nova in close binaries,
after V1324 Sco 2012 and V959 Mon 2012 \citep{ack14aa}, 
(2) the first example of coexistence of dust and soft X-ray emissions
in novae \citep{geh15eh}, 
(3) different behaviors in the $V$ and $y$ light curves: the $V$ magnitude
was recovered after the dust blackout, while the $y$ magnitude 
continuously decreased \citep{mun15mm}. 
We clarify these properties based on our shocked shell model. 

The different behaviors in the $V$ and $y$ light curves are the key 
in this study.  We present detailed $BVI_{\rm C}$ and $y$ band observation
of V339 Del from 1 day to 858 days after the discovery that were obtained
at Osaka Kyoiku University (OKU), which are shown in, e.g., Figure
\ref{v339_del_v_bv_ub_color_curve_no2} for the $V$ and $y$ magnitudes.
The $y$ magnitude is monotonically decreasing while the $V$ 
light curve has a dip around 65 days after the discovery.  
This property is first identified in the classical nova V339 Del and
a very important clue to the mechanism for the coexistence of dust 
and soft X-ray emissions. 

Formation of a dust shell is also closely related with a shock;
a dust shell forms just behind the shock \citep{der17ml}.
We also explain the temporal variations of the narrow band 
[\ion{O}{3}] and H$\alpha$ light curves as well as the broad band
$V$, $I_{\rm C}$, and $K_{\rm s}$ (or $K$),
and intermediate Str\"omgren $y$ band light curves of V339 Del. 

Our paper is organized as follows.  First we describe the $BVR_{\rm C}
I_{\rm C}$ and $y$ observation of V339 Del at Osaka Kyoiku University 
in Section \ref{observation}.  Overall observational properties 
are summarized in Section \ref{observation_summary}. 
Our shock model and expected emissions are presented in Section
\ref{shock_line_formation}.  Section \ref{light_curve_fitting} explains
our model $V$ light curve and a strong shock formation that was
derived from our nova model based on the optically-thick wind theory.  
Formation of a dust shell is discussed in connection to 
the depth of $V$ and $y$ dip after the dust blackout 
in Section \ref{discussion}.  Conclusions follow in Section \ref{conclusions}.
%Appendix gives model details, various methods 
%for obtaining distance modulus, extinction, and distance to a nova,
%as well as the time-stretching method for nova light curves,
%time-stretched color-magnitude diagram, maximum magnitude versus
%rate of decline relations. 

\section{Observation at OKU}
\label{observation}
\begin{figure}
%%\epsscale{0.75}
%%\epsscale{0.8}
%%\epsscale{1.0}
\epsscale{1.15}
\plotone{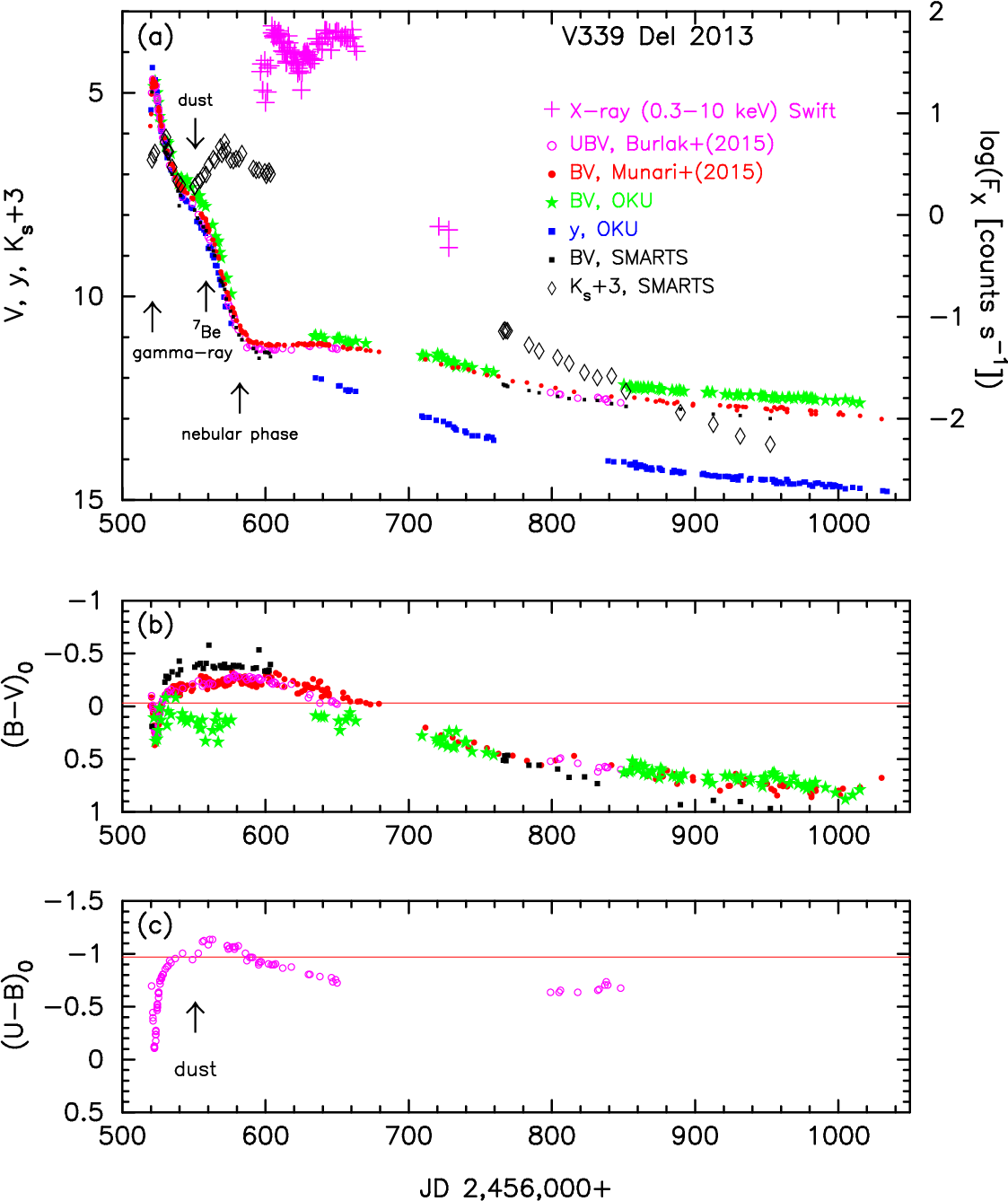}
%\plotone{v339_del_v_bv_ub_color_curve_no2.epsi}
%\plotfiddle{evolution1.ps}{5.0cm}{270}{0.4}{0.4}{-170}{220}
\caption{
The light curves and color evolutions of V339~Del on a linear timescale.
(a) The $V$, $y$, $K_{\rm s}+3$, and X-ray light curves.  
(b) The $(B-V)_0$ color curves, which are dereddened with $E(B-V)=0.18$.
(c) The $(U-B)_0$ color curve dereddened with $E(B-V)=0.18$.
The $UBV$ data are taken from \citet[][open magenta circles]{bur15a}. 
The $BV$ data are from the Small and Medium Aperture Telescope System
(SMARTS, filled black squares),
Osaka Kyoiku University (OKU) (filled green stars), and
\citet[][filled red circles]{mun15mm}.
The $y$ data are from OKU (filled blue squares).
The $K_{\rm s}$ data are from SMARTS (open black diamonds).
Dust formation was detected on JD~2,456,553 as indicated by the black arrow
labeled dust.
We add {\it Swift}/XRT data (0.3-10 keV, magenta pluses) taken
from the {\it Swift} web site \citep{eva09}.
Gamma-rays were first detected on JD~2,456,523 at the epoch denoted
by the black arrow labeled gamma-ray \citep{ack14aa}.
The first detection of $^7$Be lines on JD~2,456,585.5 \citep{taj15sn}
is denoted by the black arrow labeled $^7$Be.
\label{v339_del_v_bv_ub_color_curve_no2}}
\end{figure}

The CCD photometric observation of V339 Del at OKU was started on
UT 2013 August 14.774, approximately 4.5 hours after the discovery,
using a 0.51 m reflector telescope with optical bandpass filters
of standard Johnson-Cousins $B, V, R_{\rm C}, I_{\rm C}$, and
Str\"{o}mgren $y$.
The $y$-band filter is free from influences of striking emission
lines of novae such as \ion{Fe}{2}, [\ion{O}{3}], or [\ion{N}{2}],
so that the $y$ magnitude represents the continuum flux of the nova.
The data reduction was made by the Image Reduction and Analysis
Facility (IRAF)\footnote{\tt https://iraf-community.github.io/}
in a standard manner.  The comparison stars for the photometry were
chosen to be nearby field stars of AUIDs 000-BLC-955, 000-BLD-830,
and 000-BLF-177 from the American Association of Variable Star Observers
(AAVSO), depending on the apparent brightness of the nova
during the observing period.  Our data are shown in Figure
\ref{v339_del_v_bv_ub_color_curve_no2} together with other observational
data, the details of which will be explained in Section 
\ref{observation_summary}.

%Fig.1
%\placefigure{v339_del_v_bv_ub_color_curve_no2}

%Fig.1  

\section{Summary of observation}
\label{observation_summary}

\subsection{Overall properties of the nova light curves}
\label{observational_summary}

\citet{mun13hdc} presented $BVR_{\rm C}I_{\rm C}$ light curves during
the early 77 days, from UT 2013 August 15.115 $=$ JD 2,456,519.615
to JD 2,456,597.240.
%  This phase corresponds to the optically
%thick phase of ejecta and their transition to the nebular phase 
%\citep[2013 October 17.5 UT $=$ JD 2,456,583;][]{mun15mm}.  
They obtained the decline rates of $t_2= 10.5$ days, and $t_3= 23.5$ days
in the $V$ band, where $t_2$ and $t_3$ are the decline times by 2 and 3 mag
from the peak, respectively. 

\citet{cho14} summarized the photometric and spectroscopic evolutions
of V339~Del.  They obtained the peak brightness of $V=4.4$
on UT 2013 August 16.47 (JD 2,456,520.97), $t_2= 10$ days,
and $t_3= 18$ days for the $V$ band light curve.
They suggested the orbital period of $P_{\rm orb}= 6.43$~hr ($0.26792$~days)
from small brightness variations, and the mean values of the reddening
$E(B-V)= 0.184\pm 0.035$, peak $V$ magnitude $M_{V, \rm max}= -8.70\pm 0.03$, 
distance modulus in the $V$ band $(m-M)_V= 13.10\pm 0.08$, and the distance
$d=3.2\pm 0.3$~kpc, from various empirical laws of classical novae.
Recently, \citet{schaefer22} determined the orbital period of
$P_{\rm orb}= 0.162941\pm 0.000060$ days (3.91 hr) from TESS data.
\citet{schaefer22b} proposed a rather small distance of 
$d=1587^{+1338}_{-299}$ pc based on the Gaia eDR3 parallax.

Figure \ref{v339_del_v_bv_ub_color_curve_no2} shows the (a) $V$, $y$,
$K_{\rm s}+3$, and X-ray light curves,  and color evolutions of 
(b) $(B-V)_0$ and (c) $(U-B)_0$, on a linear timescale.
The data are taken from \citet[][$UBV$]{bur15a}, from \citet[][$BV$]{mun15mm}, 
from the Small and Medium Aperture Telescope System 
\citep[SMARTS,][$BVK_{\rm s}$]{wal12bt}, and Osaka Kyoiku University 
\citep[OKU,][$BVy$]{hac06b}.  The $y$ band is an intermediate band, the center
of which is similar to that of the $V$ band but designed to avoid strong
emission lines such as [\ion{O}{3}] lines.  Therefore, it represents 
a flux of continuum \citep[see, e.g., Figure 1 of ][]{mun13dcvf}.
Here, $(B-V)_0$ and $(U-B)_0$ are the intrinsic colors of
$B-V$ and $U-B$, which are dereddened with the color excess
of $E(B-V)=0.18$.
%%% See Appendix \ref{extinction_law}.

\subsection{Very early phase}
\label{very_early_phase}

Early photometric and spectroscopic results are reported by many groups
\citep[e.g.,][]{dar13bse, sho13sr, tom13isbd, mun13vmc, hay13cc,
shen13tt, shore13, pag13ok, mun13dc, kaw19sa}.  \citet{mun15mm} reported
their Str\"omgren $b$, $y$, and narrow-band H$\alpha$, 
[\ion{O}{3}] photometric evolution of V339 Del 
covering 500 days after the optical maximum.  

\begin{figure}
%%\epsscale{0.75}
%%\epsscale{0.8}
%%\epsscale{1.0}
\epsscale{1.15}
\plotone{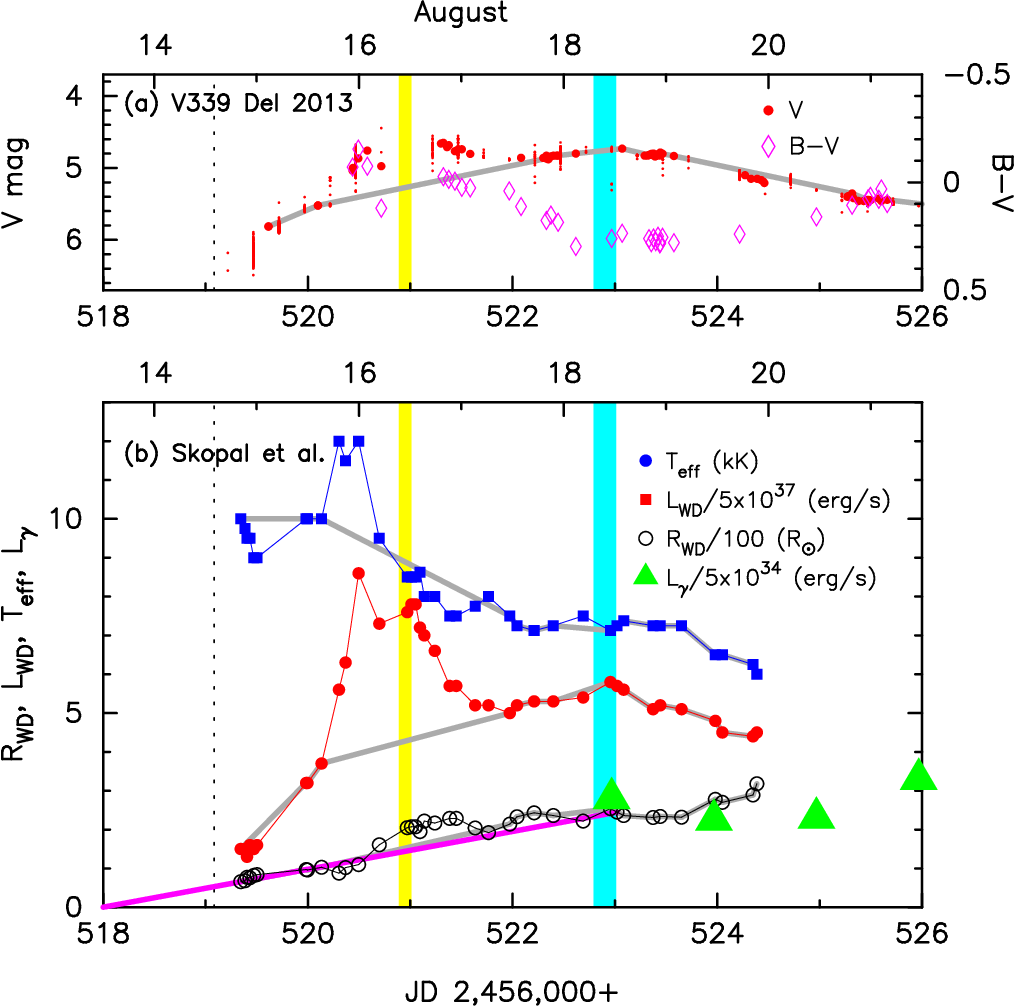}
%\plotone{v339_del_v_skopal_wd_photo.epsi}
%\plotfiddle{evolution1.ps}{5.0cm}{270}{0.4}{0.4}{-170}{220}
\caption{
(a) The early $V$ light curve (small and large filled red circles) 
and $B-V$ color evolution (open magenta diamonds) of V339~Del.
The large symbols denote the data taken from \citet{mun15mm},
\citet{bur15a}, SMARTS, and OKU, while the small filled red circles are
taken from AAVSO.  The broad gray line indicates an assumed
temporal development of the $V$ brightness if we do not include the early
fluctuations of the $V$ data.
The vertical dashed line indicates the epoch of discovery by Itagaki while
the vertical broad yellow and cyan lines correspond
to the epochs of the first and second optical peaks, respectively, which are
taken from \citet{sko14dt}.
(b) The temporal developments of the effective temperature $T_{\rm eff}$,
luminosity $L_{\rm WD}$, and radius $R_{\rm WD}$ of the pseudo-photosphere,
taken from \citet{sko14dt}.  The luminosity and radius depend on
the assumed distance to the nova.  \citet{sko14dt} assumed $d=3$~kpc,
so that the luminosity and radius change according to 
$L_{\rm WD} \propto (d/3{\rm ~kpc})^2$ and
$R_{\rm WD} \propto (d/3{\rm ~kpc})$, respectively.
The three thin blue, red, black lines connect each data.
We also add GeV gamma-ray fluxes \citep[filled green triangles;][]{ack14aa}.
\citet{ack14aa} assumed $d=4.2$ kpc, so the
gamma-ray luminosity depends on $L_{\gamma} \propto (d/4.2{\rm ~kpc})^2$.
\label{v339_del_v_skopal_wd_photo}}
\end{figure}

Figure \ref{v339_del_v_skopal_wd_photo} shows the
(a) early $V$ light curve and $B-V$ color evolution of V339~Del and 
(b) temporal variations of the luminosity, radius, and effective
temperature of the white dwarf (WD) pseudo-photosphere obtained
by \citet{sko14dt}.  Note that the luminosity and radius depend on
the assumed distance to the nova (see figure captions).  
We also added the gamma-ray fluxes reported by \citet{ack14aa}.

%Fig.2
%\placefigure{v339_del_v_skopal_wd_photo}

%Fig.2  

\subsection{Expansion of the photosphere, optical peak, and outburst day}
\label{photosphere_peak_outburstday}

The $V$ light curve in Figure \ref{v339_del_v_skopal_wd_photo} 
shows the two maxima. The first peak is $V=4.4$ 
on UT 2013 August 16.47 $=$ JD 2,456,520.97 \citep{cho14}, 
as indicated with the vertical broad yellow line. 
The data of the yellow line are taken from \citet{sko14dt}, i.e., 
UT 2013 August $16.45\pm0.06$.

The second $V$ peak is $V=4.7$ on UT 2013 August 18.4$\pm 0.11 =$
JD 2,456,522.9$\pm 0.11$ \citep{sko14dt}, as indicated with 
the vertical broad cyan line.  The $B-V$ color evolution shows that
this second maximum corresponds to a global maximum of $B-V$ color
(see Figure \ref{v339_del_v_skopal_wd_photo}a).
Thus, we regard this second $V$ maximum to be a global maximum without
the effect of local (short timescale) fluctuations like in the first peak.

We extrapolated the photospheric radii in 
Figure \ref{v339_del_v_skopal_wd_photo}(b)
to zero, i.e., $t_{\rm OB}=$ JD~2,456,518.0
(UT 2013 August 13.5) as indicated by the straight magenta line. 
\citet{schaefer14bg} estimated the outburst day to be
JD~2,456,518.077 (UT 2013 August 13.577) by extrapolating the two 
pre-discovery magnitudes to the quiescent brightness. 
Our estimated date is consistent with the Schaefer et al.'s date.
In what follows, we define time $t$ as the days after outburst, i.e.,
time from $t_{\rm OB}$.

The outburst day could be exactly identified 
if the X-ray flash was detected in V339 Del. 
The X-ray flash is a brief X-ray bright phase just 
after the onset of nova outburst,   
several days before optical brightening \citep{kat22shb,kat22shc}. 
The classical nova YZ Ret is only the nova in which 
the X-ray flash is detected \citep{kon22wa}.

\subsection{Distance}
\label{distance}

\citet{sho13sa} and \citet{shore13} obtained the distance of $d=4.2$~kpc
comparing the UV fluxes of V339~Del with those of OS~And \citep{sho16ms}. 
This distance of $d=4.2$~kpc, however, is much larger than $d=2.06^{+1.22}
_{-0.75}$~kpc of Gaia early data release 3 (Gaia eDR3) \citep{bai21rf}.  
Shore et al. assumed $E(B-V)=0.25\pm0.05$ and $d=5.1$~kpc for OS~And.
If we adopt $d=3.6^{+1.8}_{-1.0}$~kpc for OS And
from the Gaia eDR3 distance \citep{bai21rf} and assume that the UV flux
of V339 Del should be 1.25 times larger than that of OS And 
\citep[e.g.,][]{hac16kb} because
the WD mass of V339 Del is about 1.25 times more massive than that of OS And,
we obtained a much smaller distance of
$d\sim 4.2\times (3.6/5.1)/\sqrt{1.25}= 2.5$ kpc to V339 Del. 
%This value is much closer to the Gaia eDR3 distance than
%\citet{sho16ms}'s $d=4.2$ kpc.

\citet{cho14} gave a distance of $d=3.2\pm 0.3$~kpc from various empirical
laws on the absolute magnitudes of classical novae.
These empirical laws such as the maximum magnitude versus rate of
decline (MMRD) relation, however, have a large scatter for an individual nova 
as shown later in Figure \ref{vmax_t3_vmax_t2_selvelli2019_schaefer2018_2fig}
\citep[see also ][for discussion]{schaefer18}.
Therefore, we do not expect a high accuracy for the distance.

\citet{schaefer14bg} obtained the distance to the nova to be 
$d=4.54 \pm 0.59$ kpc from the expansion parallax method
based on their interferometric image data in the nova fireball stage,
together with the expansion velocity of 
the ejecta, $613 \pm 79$ km s$^{-1}$.

The expansion rate of the photosphere in V339 Del can be estimated from 
Figure \ref{v339_del_v_skopal_wd_photo}(b), which shows the temporal
variation of the pseudo-photosphere \citep{sko14dt}.
The mean expansion rate (magenta line)
of the pseudo-photosphere is approximately
$d R_{\rm ph}/dt = 4\times 10^7$~cm~s$^{-1}~(d/3~{\rm kpc})$ = 400~km
s$^{-1}~(d/3~{\rm kpc})$ from UT August 14.84 to the second $V$ peak 
(UT August 18.4).  If we use the distance of $d=2.06^{+1.22}_{-0.75}$~kpc
\citep{bai21rf}, this photosphere expansion rate
is about $d R_{\rm ph}/d t=$280 km s$^{-1}$, which is much smaller than
the wind velocity of $v_{\rm wind}=613 \pm 79$ km s$^{-1}$
\citep{schaefer14bg}.

Thus, once winds begin to blow, the photosphere expansion rate
cannot be determined from the spectra, because the winds leave
the photosphere behind.  This is a reason why Schaefer et al.
failed to obtain the correct distance.  If we adopt $d R_{\rm ph}/d t
=$280 km s$^{-1}$, we obtain the distance of $\sim 2.06$ kpc.
(This is a kind of tautology, however.  Assuming $d=2.06$ kpc, we obtain
$d=2.06$ kpc.)  

\citet{geh15eh} also estimated the distance to the nova to be 
$4.4953 \pm 0.8$ kpc from the expansion rate of the photosphere
with their Equation (3) and the expansion velocity of the ejecta,
505 km s$^{-1}$.  If we instead adopt $d R_{\rm ph}/dt=$280 km s$^{-1}$,
we similarly obtain the distance to the nova to be $\sim 2.5$ kpc.

These examples give us a warning that the expansion parallax method
cannot be applied to a nova fireball stage, simply because the expansion
rate of the pseudo-photosphere is different from the velocity of ejecta
\citep[see, e.g.,][for nova winds]{kat22sha}.

In the present paper, we determine the distance 
to V339 Del to be $d=2.1\pm 0.2$~kpc for
the extinction $E(B-V)= 0.18$ \citep{mun13vmc, cho14}
by using the time-stretching method \citep{hac10k, hac15k, hac16k,
hac18k,hac20skhs}, which will be described in Section
\ref{v339_del_time_stretch}.

\subsection{E(B-V)}
\label{extinction_distance}

\citet{rud13} reported on $0.45 - 2.5\mu$m spectroscopy of V339 Del.
The object is an ``\ion{Fe}{2}''-type nova\footnote{\citet{wil92}
divided novae into two major classes depending on their early spectra.
Either ``\ion{Fe}{2} lines'' or ``emission lines of He or N'' is stronger.}  
with numerous infrared \ion{Fe}{2} features.
In addition, there is strong emission from \ion{C}{1}, \ion{N}{1},
\ion{O}{1}, and \ion{Ca}{2}.  The [\ion{O }{1}] lines are present,
including the line at 557.7 nm.  Line widths for the stronger hydrogen
lines are approximately 1300 km s$^{-1}$ (FWHM).
The interstellar reddening, as measured from the \ion{O}{1} lines,
is $E(B-V) = 0.33 \pm 0.1$. 

\citet{nel13} obtained the  interstellar absorption column density of
$N_{\rm H}\sim 1.8\times 10^{21}$~cm$^{-2}$ from the Chandra X-ray
spectrum, 88.25 days after the outburst.
This column density corresponds to 
the reddening of $E(B-V)= N_{\rm H}/8.3\times 10^{21}
=1.8\times 10^{21}/ 8.3\times 10^{21}= 0.22$
based on the relation given by \citet{lis14}.

\citet{tom13isbd} estimated the reddening of $E(B-V)\sim 0.17$ from the
equivalent width of interstellar \ion{Na}{1} D1 line and the relation
given by \citet{mun97z}.
\citet{mun13vmc} also estimated the reddening of $E(B-V)= 0.182$ from the
equivalent width of interstellar \ion{Na}{1} D1 line and the relation
given by \citet{mun97z}.
\citet{cho14} obtained a similar value of $E(B-V)= 0.184\pm 0.035$
from the mean value calculated with various methods as mentioned in 
Section \ref{observational_summary}.

We plot the $(B-V)_0$ and $(U-B)_0$ colors dereddened with $E(B-V)= 0.18$
in Figure \ref{v339_del_v_bv_ub_color_curve_no2}b and c.
The $(B-V)_0$ colors diverge between JD 2,456,530 and JD 2,456,710 
among the different telescopes.  In general, this kind of divergence
frequently occurs, when strong emission lines contribute to each band edge.
If the $V$ filter response is slightly different in each telescope,
strong emission lines such as [\ion{O}{3}] on the blue edge of 
$V$ filter make a large difference in the flux.  Such examples and
discussion can be seen, e.g., in \citet{mun13dcvf}.

In the present paper, we adopt the extinction of $E(B-V)= 0.18$
after \citet{tom13isbd}, \citet{mun13vmc}, and \citet{cho14},
then determine the distance modulus in the $V$ band to be
$(m-M)_V=12.2\pm 0.2$ by using the time-stretching method
\citep{hac10k, hac15k, hac16k, hac18k,hac20skhs},
which will be described in Section \ref{v339_del_time_stretch},
and obtain the distance to be $d=2.1\pm 0.2$~kpc.

\subsection{Gamma-ray emission}
\label{gamma_ray_flux}

\citet{hay13cc} reported the detection of GeV gamma-rays from V339~Del
at $>5 ~\sigma$ significance on UT 2013 August 18.5 (day 5).
%%%(JD 2,456,522.5). >>> 523.0-518.0=5
This epoch is indicated with the upward arrow 
in Figure \ref{v339_del_v_bv_ub_color_curve_no2}a.

\citet{ack14aa} reported the detail of the Fermi/LAT observation on V339 Del.
We added the early GeV gamma-ray fluxes in Figure
\ref{v339_del_v_skopal_wd_photo}b. 
The peak flux and the total energy were estimated to be $L_\gamma\sim
2.6\times 10^{35}$ erg s$^{-1} ~(d/4.2~{\rm kpc})^2$  and
$E_\gamma\sim 6\times 10^{41}$~erg $~(d/4.2~{\rm kpc})^2$, respectively.

\subsection{IR light curve and dust formation}
\label{infrared_dust}

Figure \ref{v339_del_v_bv_ub_color_curve_no2}a depicts the $K_{\rm s}$
light curve, which shows two maxima.
% that shows a wavy structure. 
We will discuss this behavior of the $K_{\rm s}$ light curve in
relation to our model later in Section \ref{thin_dust_shell}. 

\citet{shen13tt} reported the formation of warm dust ($T_{\rm d}\sim 1000$~K) 
on UT 2013 September 21.8 (day 39.3) and September 27.8 (day 45.3).
%%% (JD 2,456,557.3 and JD 2,456,563.3), 
%%%$\sim 45$ and $\sim 51$ days after the outburst. 
%%% K-L=1.35, L-M=0.42 on 21.8
%%% K-L=1.59, L-M=0.45 on 27.8
The increase of brightness in $K_{\rm s}$ or
$KLM$ bands occurred on UT 2013 September 17.0 (day 34.5) 
%%%JD~2456552.5 => 552.5 - 518= 34.5
%%%JD~2,456,518.0 =
in the data of SMARTS and \citet{bur15a}.
\citet{eva17bg} obtained the formation date of dust to be 34.75 days after
the eruption (their $t=0$ is UT 2013 August 13.9 $=$ JD~2,456,518.4).

The epoch of dust formation is indicated by the black downward arrow labeled 
dust in Figure \ref{v339_del_v_bv_ub_color_curve_no2}a, $\sim 35$ days
after the outburst.
%%% ($t_{\rm OB}=$ UT 2013 August 13.5 = JD~2,456,518.0).  

%Fig.3ab
\begin{figure*}
%%\epsscale{0.75}
%%\epsscale{0.8}
\epsscale{1.0}
%%\epsscale{1.15}
%%\plotone{f9.eps}
\plottwo{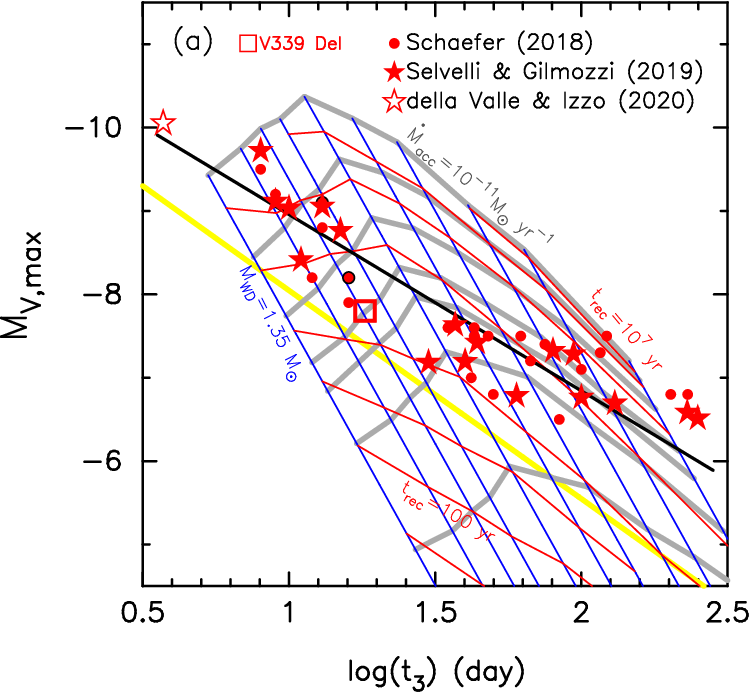}{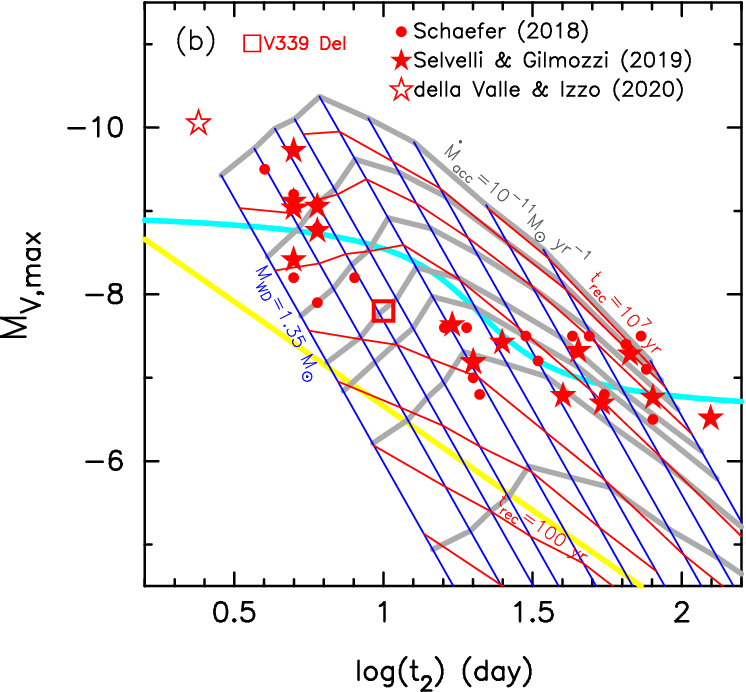}
%\plotone{v1668_cyg/outburst/max_t3_v339_del_schaefer2018_all_saio_kato2014_no10.epsi}
%\plotone{v1668_cyg/outburst/max_t2_selvelli2019_schaefer2018_all_saio_kato2014_no9.epsi}
%\plotfiddle{evolution1.ps}{5.0cm}{270}{0.4}{0.4}{-170}{220}
\caption{
The maximum $V$ magnitude against the rate of decline, (a) $t_3$ and
(b) $t_2$, for V339~Del (open red square), as well as for other
classical novae.  The original data are taken from Figure 6
of \citet{hac20skhs}.  The filled red circles represent novae taken from
``Golden sample'' of \citet{schaefer18}, from which we exclude recurrent
novae and V1330~Cyg.  The filled stars are novae taken from
\citet{sel19}.  The open star is V1500~Cyg taken from \citet{del20}.
The blue straight lines indicate model equi-WD mass lines, from left to right,
1.35, 1.3, 1.25, 1.2, 1.1, 1.0, 0.9, 0.8, 0.7, and $0.6~M_\sun$;   
the thick solid gray lines denote model equi-mass accretion rate 
($\dot M_{\rm acc}$) lines, from lower to upper, $3\times 10^{-8}$,
$1\times 10^{-8}$, $5\times 10^{-9}$, $3\times 10^{-9}$, $1\times 10^{-9}$, 
$1\times 10^{-10}$, and $1\times 10^{-11} M_\sun$~yr$^{-1}$;
the red lines represent model equi-recurrence time lines, from lower to upper,
$t_{\rm rec}= 30$, 100, 300, 1000, 10000, $10^5$, $10^6$, and $10^7$~yr.
These lines are taken from \citet{hac20skhs} based on the universal
decline law of novae and model calculation of mass accretion onto each WD. 
The thick yellow line corresponds to the $x_0=2$ line, below which
the models are not valid \citep[see ][for details]{hac20skhs}.
In panel (a), the thick solid black line indicates the empirical
line for the MMRD relation obtained by \citet{sel19}.
In panel (b), the thick solid cyan line represent the empirical MMRD
line obtained by \citet{del20}.
\label{vmax_t3_vmax_t2_selvelli2019_schaefer2018_2fig}}
\end{figure*}

\subsection{Supersoft X-ray source phase}
\label{supersoft_x-ray_source}

The bright supersoft X-ray source (SSS) phase started 
on UT 2013 October 24 \citep[day 71.5;][]{osb13pb}.
%%%[JD~2456589.5,][]
%$\sim 72$ days after the outburst. 

Note that this epoch corresponds broadly to the time that 
the nova V339~Del entered the nebular phase
on UT 2013 October 18 (day 65.5)
%%%(JD 2,456,583.5)
from the flux ratio of [\ion{O}{3}]$/$H$\beta >1$ \citep{mun13dc}.
\citet{mun15mm} pointed out that the [\ion{O}{3}] 5007 \AA\ emission line
rapidly grows in intensity, because the spreading ionization is caused
by the supersoft X-ray input.  

\citet{nel13} reported a Chandra X-ray spectrum of V339~Del 
during the SSS phase on UT 2013 November 9.75 (day 88.25).
%%%= JD 2,456,606.25
%%%$\sim 89$ days after the outburst.
The spectrum shows many absorption lines superimposed on a
supersoft continuum source.  Their blackbody fit with the continuum
indicates a temperature of k$T\sim 27$~eV ($T\sim 3.1\times 10^5$~K) 
and interstellar absorption column density of 
$N_{\rm H}\sim 1.8\times 10^{21}$~cm$^{-2}$.

\citet{pag14ko} reported that the SSS phase has ended 
before UT 2014 March 4 (day 202.5).
%%%, $\sim 206$ days after the outburst.
%%%  (JD~2456720.5)    720.5 - 518.0 = 202.5
It is difficult to determine the exact turn-off date 
because there is a large gap between the last observation
before the Swift Sun constraint (UT 2014 January 6= day 145.5)
and UT 2014 March 4 (day 202.5). 
% JD 2456663.5 >>> 663.5-518=145.5
% JD 2456720.5 >>> 720.5-518=202.5

\citet{bea13op} reported a $54$~s quasi-periodic oscillation
with Swift X-ray observation.  \citet{nes13}
also confirmed the 54~s oscillation with XMM-Newton.

\citet{osb13pb} suggested that the start of the SSS phase 
coincides with the appearance of the plateau phase in the $V$ band.
%%% 589.5 - 518.0 = 71.5
Such coincidence is observed in recurrent novae like U~Sco, which 
has been theoretically explained as effects of 
a large irradiated accretion disk \citep[e.g.,][]{hkkm00}. 
In V339 Del, however, Osborn et al.'s explanation is unlikely because 
(1) the $y$ magnitude light curve does not show a flat plateau 
(Figure \ref{v339_del_v_bv_ub_color_curve_no2}a) 
and  (2) the short orbital period of $P_{\rm orb}= 3.91$~hr ($=0.163$ days)
\citep{schaefer22} hardly supports a bright (large) accretion disk
like in U Sco whose orbital period is $P_{\rm orb}= 29.5$~hr ($=1.23$ days)
\citep{sch95r}.

\subsection{Nova speed class ($t_2$ and $t_3$) versus peak $V$ brightness}
\label{mmrd_relation}

\citet{hac20skhs} calculated a number of nova model light curves and 
constructed theoretical diagrams for the relations among various nova
parameters.  Figure \ref{vmax_t3_vmax_t2_selvelli2019_schaefer2018_2fig}
shows such diagrams, the so-called maximum magnitude versus rate of decline
(MMRD) diagrams for novae.  If the position of a nova is given in
these MMRD diagrams, we can know its WD mass, mass-accretion rate,
and recurrence time.  For V339 Del, the rates of decline, $t_3= 18$
and $t_2= 10$ days, were obtained by \citet{cho14}.  The absolute maximum
$V$ magnitude is calculated to be $M_{V,\rm max}= m_{V,\rm max} - (m-M)_V
= 4.4 - 12.2 = -7.8$.  We plot these two positions in  
Figure \ref{vmax_t3_vmax_t2_selvelli2019_schaefer2018_2fig}a and b,
respectively.  Both the positions are located on the blue line
of $M_{\rm WD}= 1.25~M_\sun$, and on the gray line of 
$\dot M_{\rm acc}=3\times 10^{-9}~M_\sun$~yr$^{-1}$.
The recurrence time can be estimated to be
$t_{\rm rec}\sim 2,000$~yr from an interpolation between the two red lines
of $t_{\rm rec}\sim 1,000$~yr and $t_{\rm rec}\sim 10,000$~yr.   
In this way, these theoretical diagrams are useful to easily
estimate the nova parameters such as the WD mass $M_{\rm WD}$,
average mass-accretion rate to the WD $\dot{M}_{\rm acc}$,
and recurrence time $t_{\rm rec}$ from the decline rates and
peak $V$ brightness.

The above results are consistent with our $1.25 ~M_\sun$ WD model 
obtained in Section \ref{light_curve_fitting}.

\begin{figure*}
\gridline{\fig{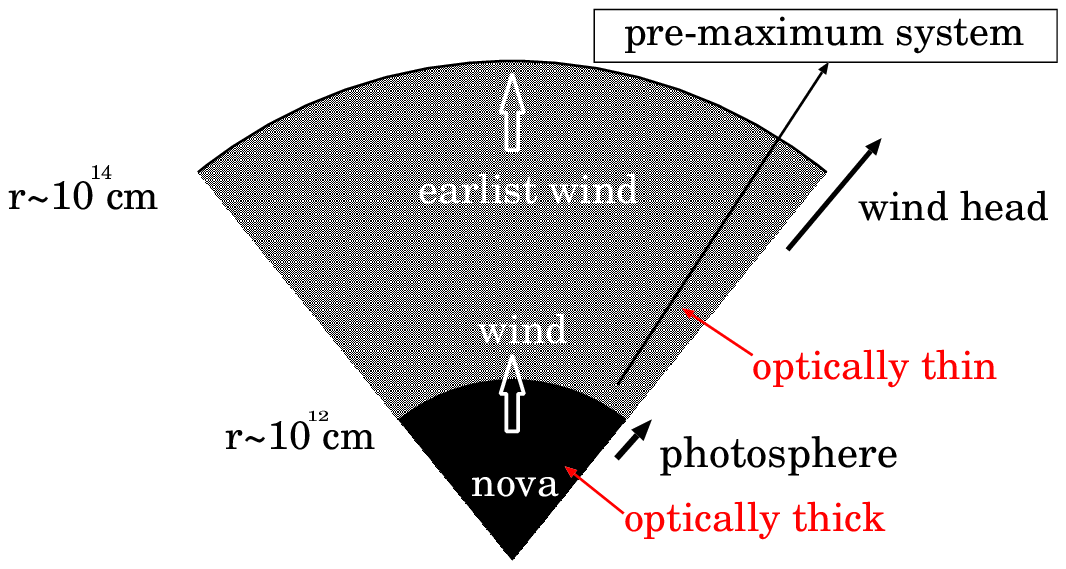}{0.22\textwidth}{(a) pre-maximum phase}
          \fig{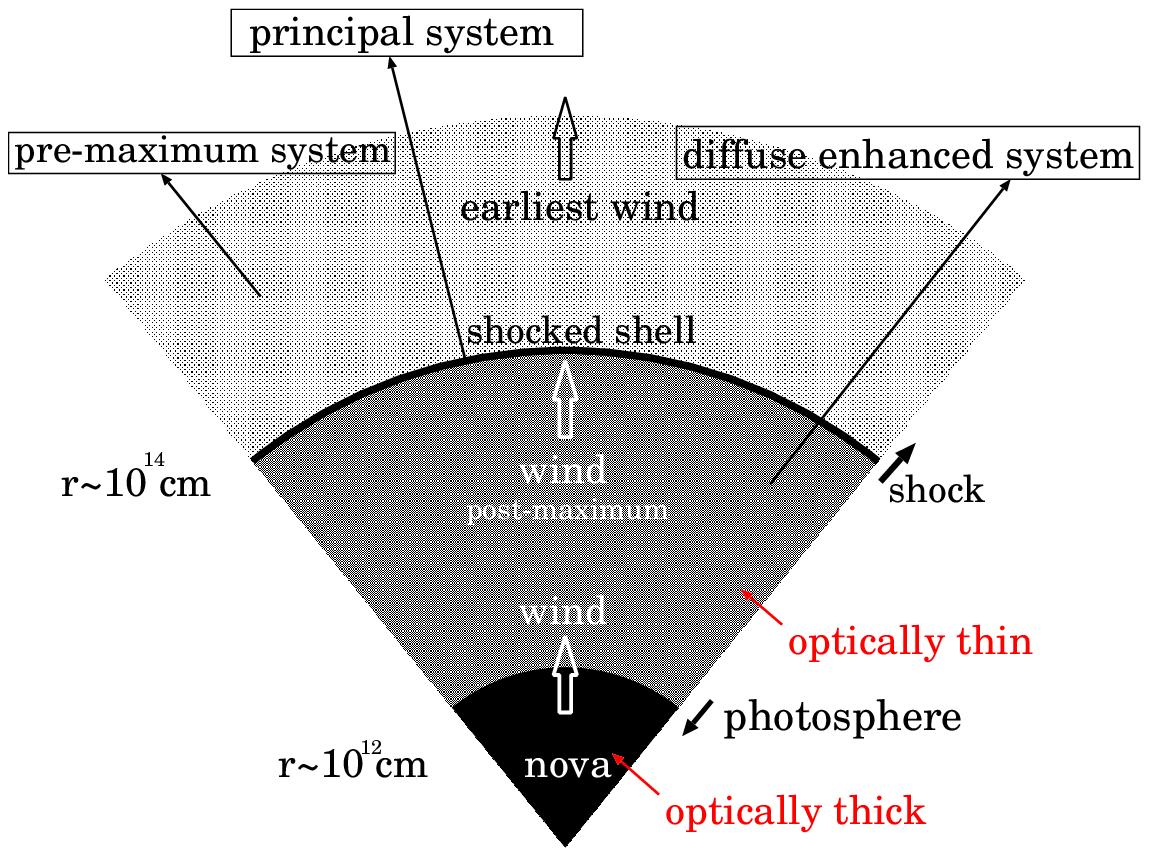}{0.35\textwidth}{(b) post-maximum phase}
          \fig{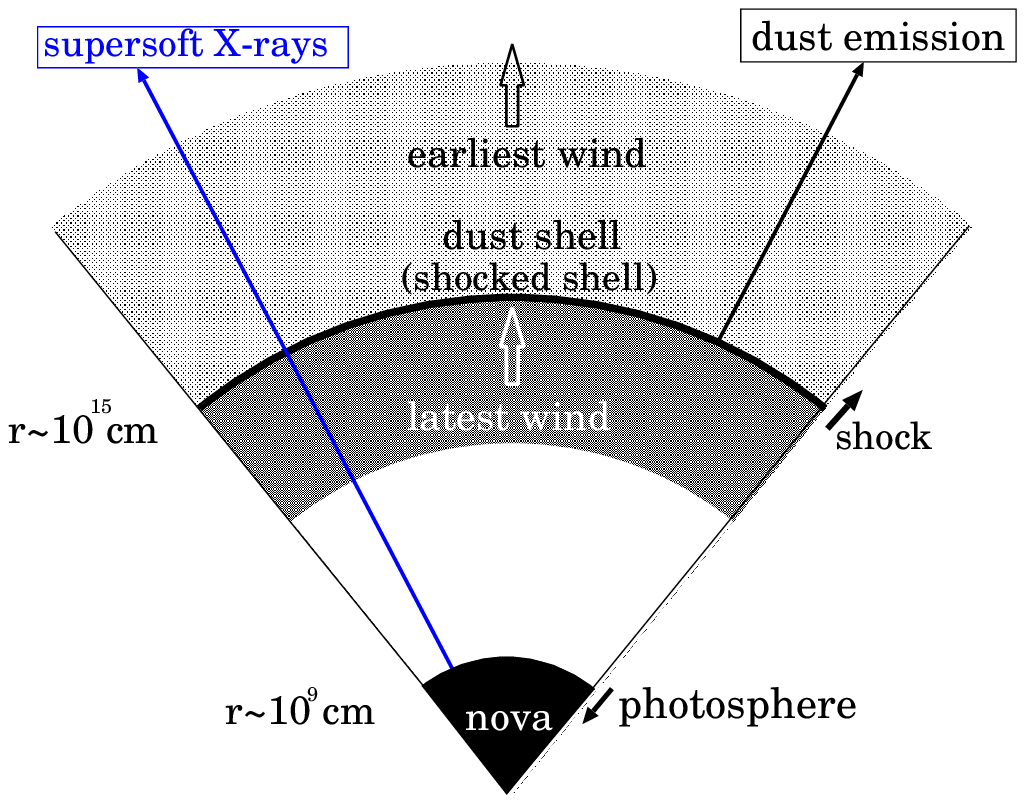}{0.4\textwidth}{(c) supersoft X-ray source phase}
          }
\gridline{
          \fig{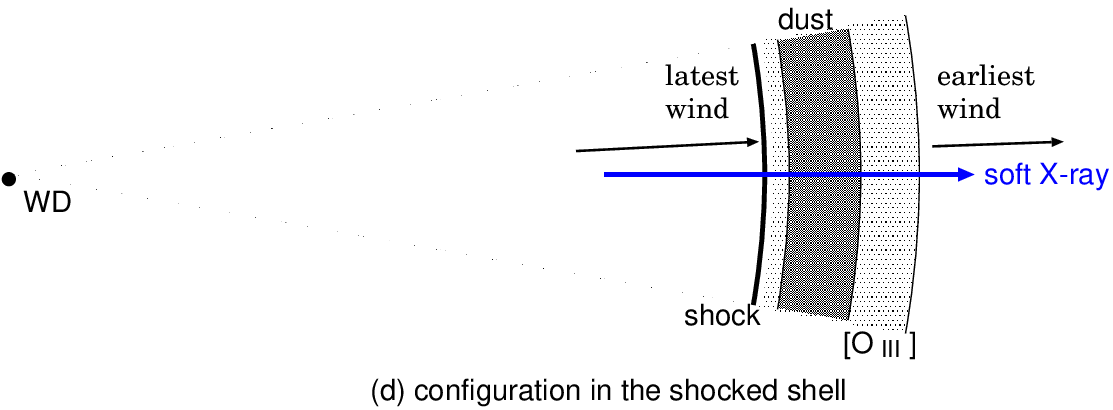}{0.75\textwidth}{}
%          \fig{f2d.eps}{0.45\textwidth}{}
          }
%\begin{figure}
%\epsscale{1.15}
%%%\plotone{f1.eps}
%\plottwo{f1.eps}{f2.eps}
%%%%\plotone{mdot_radius_velocity.epsi}
%%%%\plotone{opt_gamma_ray_fluxes.epsi}
%\plotfiddle{evolution1.ps}{5.0cm}{270}{0.4}{0.4}{-170}{220}
\caption{
Cartoon for our V339 Del nova model.  
(a) The nova photosphere expands over $R_{\rm ph} \sim 0.1 ~R_\sun$ and 
optically thick winds are accelerated deep inside the photosphere
\citep{kat22sha}. 
The wind itself becomes optically thin outside the photosphere.
The earliest wind forms the pre-maximum absorption/emission line system
\citep{mcl42} outside the photosphere ($r > R_{\rm ph}$).
(b) After optical maximum, the photosphere is receding and
a strong shock is formed outside the photosphere \citep{hac22k}.
The shocked layer is geometrically thin, and the whole ejecta is
divided into three parts, outermost expanding gas (earliest wind),
shocked shell, and inner wind.  These three parts contribute to 
pre-maximum, principal, and diffuse enhanced absorption/emission line
systems \citep{mcl42}, respectively, as proposed by \citet{hac22k, hac23k}.
The shocked shell emits thermal hard X-rays.
(c) The optically thick wind has stopped.  The photospheric temperature
becomes high enough to emit supersoft X-rays.  The latest wind still 
collides with the shocked shell.  A dust shell forms behind the shock
and contributes to dust emission.
(d) An enlargement of the shocked layer in panel c.  We plot
locations of shock, dust shell, and [\ion{O}{3}] emission region.
\label{wind_shock_config}}
\end{figure*}

\section{Shock formation}
\label{shock_line_formation}

As described in the previous section, V339 Del was well observed in
multiwavelength bands.  Such a large variety in wavelengths
(or energies) of emissions indicates that various emissions originate from
different places or times of the nova ejecta. 
\citet{hac22k, hac23k} %and \citet{hac23k}
proposed nova ejecta evolution models 
based on the fully self-consistent nova outburst model of \citet{kat22sha}. 
The authors found that a strong shock naturally arises in nova ejecta
far outside the WD photosphere without ad hoc or arbitrary setting on 
the nova ejecta, and elucidated the origin of nova
absorption/emission line systems raised by \citet{mcl42}.

Optically thick winds start after the X-ray flash ends \citep{kat22shb,
kon22wa} and continue to blow until about the beginning of the SSS phase. 
The winds are called ``optically thick winds,'' because matter is
accelerated deep inside the photosphere \citep{bat78, rug79b, kat94h}.
However, the wind itself
becomes optically thin outside the photosphere as shown in Figure
\ref{wind_shock_config}a and b.
The optical flux is dominated by free-free emission from ejecta 
just outside the photosphere of a nova envelope. 
The flux of free-free emission is larger for a larger mass-loss 
rate of the wind.  In \citet{kat22sha}'s model, when the wind mass-loss
rate reaches maximum, both the envelope photospheric
radius and free-free emission brightness also reach maximum
\citep[see Figure 2 of ][]{hac22k}.

The wind velocity at the photosphere becomes slower toward the optical
maximum \citep{kat22sha, hac22k}.  Such a tendency of decreasing wind
velocity toward optical maximum was observed in many novae 
\citep[e.g., ][]{ayd20ci}.
This simply means that the wind ejected earlier leaves behind the wind
ejected later and therefore wind fluid expands.  After the optical maximum, 
the wind velocity at the photosphere turns to increase.  The wind
ejected later can catch up the wind ejected earlier and
compression of wind fluid occurs, resulting in
a formation of a strong shock, as shown in Figure
\ref{wind_shock_config}b.

Gamma-rays and hard X-rays are originated from the shock
which is located far outside the WD photosphere.
%%%These happen after the optical maximum.  
Supersoft X-rays are emitted from the WD photosphere 
when the photospheric temperature becomes high enough
near/after optically-thick winds stop (Figure
\ref{wind_shock_config}c). 

The characteristic properties of nova winds play essential roles in
the configuration of nova ejecta and emissions with different 
energy/wavelength.
Figure \ref{wind_shock_config} illustrates such a nova ejecta model 
for V339 Del.  In what follows, we study observational properties
of V339 Del along with this nova ejecta evolution scenario.

\subsection{Very early phase before optical maximum}
\label{pre-maximum_phase}

Figure \ref{wind_shock_config}a illustrates a very early phase
of the nova outburst.  The hydrogen-rich envelope of the WD begins
to expand and optically thick winds start to blow
when the photospheric temperature decreases to 
$\log T_{\rm ph}$ (K)$\approx 5.3$ and the photospheric radius
expands to $R_{\rm ph} \sim 0.1 ~R_\sun$.
After that, the photospheric radius increases with time until 
the maximum expansion of the photosphere. 
We call this period the pre-maximum phase.

In \citet{kat22sha}'s nova model, the photospheric velocity of wind
is decreasing with time until the optical maximum (or maximum expansion
of the photosphere).  Such a decreasing trend of velocity is commonly
observed among various novae \citep[e.g.,][]{ayd20ci}.

The same trend of decreasing velocity is also obtained in V339 Del 
before the global optical $V$ maximum. 
\citet{sko14dt} estimated the wind velocity
from the H$\alpha$ P Cygni absorption line center. 
The velocity is decreasing from $-1600$ km s$^{-1}$
(on $t=1.5$ days) to $-900$ km s$^{-1}$ (on $t=5$ days $\approx$ the epoch of
global optical $V$ maximum). 
\citet{tom13isbd} also obtained $-1600$ km s$^{-1}$ 
(JD 2,456,519.38 $=$ UT 2013 August 14.88), 1.38 days after the outburst. 
A similar trend of decreasing velocity in the pre-maximum phase was
also reported by \citet{deg15sm, deg16sm}.

It should be noted that the expansion rate of the pseudo-photosphere,
$d R_{\rm ph}/d t$, where $R_{\rm ph}$ is the photospheric radius,
is not the same as the photospheric wind velocities, $v_{\rm ph}$ or
$v_{\rm wind}$, as illustrated in Figure \ref{wind_shock_config}a and
already mentioned in Section \ref{distance}.
After optical maximum, the photosphere begins to shrink, not expands,
as illustrated in Figure \ref{wind_shock_config}b.
 
%%
%% ~250 R_sun  in 5 days  
%% 250 x 7E10 / 5 x 24 x 3600 = 405 km/s
%%
%%(1/2)v^2=GM/R    v=(2GM/R)^{1/2} = (2 x 6.67E-8 x 2E33 / 250 x 7E10)^{1/2}
%%  =  39 km/s (d/3 kpc)^{-1/2}
%%
%Then the expansion velocity of the pseudo-photosphere is about
%$v_{\rm exp}\sim 400$~km~s$^{-1}~(d/3~{\rm kpc})$, which is much faster
%than the escape velocity $v_{\rm esc}\sim 40$~km~s$^{-1}~(d/3~{\rm kpc})^{-
%1/2}$ from a $\sim 1.0~M_\sun$ WD at the
%pseudo-photosphere ($R_{\rm ph}\sim 250 ~R_\sun~(d/3~{\rm kpc})$)
%on JD~2,456,523 ($t=5$ days).

\subsection{Post-maximum phase and shock formation}
\label{post_maximum_phase}

\citet{hac22k} found a formation of a strong shock based on 
\citet{kat22sha}'s fully self-consistent nova outburst model.  
In their nova model, the wind velocity is decreasing with time toward 
the optical maximum, but turn to increase after that.  
The wind trajectories depart from one another before the optical maximum
(expansion), but converge after the optical maximum (compression). 
In other words, after the optical maximum, the wind ejected later
catches up with the wind ejected earlier.  This makes a reverse shock. 
Thus, a strong shock arises far outside the photosphere.
This shock formation mechanism is common among all the WD masses and
mass-accretion rates as far as the main driving force of wind
is the radiative opacity \citep{hac22k, kat22sha}.

The trend of increasing wind velocity after optical maximum
can be explained as follows: 
In optically-thick winds, matter is accelerated deep inside the photosphere
by radiation pressure gradients which depend on the OPAL opacity 
\citep{igl96r}.  
%The WD envelope expands to emit strong winds after the nova went into 
%outburst. The winds are called ``optically-thick winds,'' because matter is
%accelerated deep inside the photosphere \citep{bat78, rug79b, kat94h}.
%It is obvious, from the definition of photosphere, that the wind itself
%becomes optically thin outside the photosphere
%(Figure \ref{wind_shock_config}a).
%Once optically-thick winds begin to blow in
%an early phase of the nova outburst, the wind velocity is not the same
%as the expansion rate of the photosphere \citep[e.g.,][]{kat22sha}, that is,
%$v_{\rm wind} > d R_{\rm ph}/d t$, where $v_{\rm wind}$ is the velocity
%of wind and $R_{\rm ph}$ is the radius of the photosphere, because,
%in general, winds leave the photosphere behind. 
In the decay phase of a nova, the mass of hydrogen-rich
envelope is rapidly decreasing due to wind mass-loss.
%  If the radiation
%pressure gradients are the same, the terminal wind velocity is
%increasing with time because the envelope mass above the critical point
%of wind is quickly decreasing due to wind mass-loss.
%In other words, 
A less massive envelope is accelerated to a higher velocity
for a given momentum (same radiation pressure gradients). 
%Note, however, that this is a simplest explanation, neglecting other factors.

In this post-maximum phase, the whole ejecta is divided into three parts, 
as illustrated in Figure \ref{wind_shock_config}b;
the outermost expanding gas (earliest wind),
shocked shell (geometrically thin), and inner wind. 
These three parts contribute to pre-maximum, principal, 
and diffuse enhanced absorption/emission line systems \citep{mcl42},
respectively, as proposed by \citet{hac22k, hac23k}.
It should be stressed that this interpretation is straightforward because
we do not need any other arbitrary assumption (or setting) on nova ejecta.

In the V339 Del outburst, the wind velocity is obtained as follows. 
After the optical maximum ($t=5$ days), the bluest edge of P Cygni
absorption increases to $-1300$ km s$^{-1}$ on day 16--26 
\citep[see, e.g., Figure 8 of ][]{deg15sm}.  \citet{sho16ms} obtained
$-1400$ km s$^{-1}$ on day 16-37 (see their Figure 2).
We regard these velocities
to be the diffuse-enhanced absorption system.  On the other hand,
the main absorption part is located at $-800$ km s$^{-1}$ \citep{deg15sm}
or at $-900$ km s$^{-1}$ \citep{sho16ms} both on the same days mentioned
above.  We regard these to be the principal absorption system.

\subsection{GeV gamma-rays from the shocked shell} 
\label{first_detection_gamma-rays}

Gamma-rays have been detected in $\sim 20$ novae \citep[e.g.,][]{cho21ms};
some of them are in symbiotic binaries, but many are in close binaries  
that has no massive circumbinary matter.  The so-called internal shock
has long been expected in close binary novae as an origin of
high energy (hard) X-rays \citep[e.g.,][]{fri87, muk01i}.  Here,
the internal shock simply means the shock in the ejecta, 
and the external shock means the shock between the circumstellar
matter and the nova ejecta.

\citet{hac22k} theoretically found that a shock wave inevitably arises
in the course of nova outburst without any arbitrary setting, 
such as multiple shells or a circumbinary matter. 
They further clarified that a shock forms only after the
optical maximum.

\citet{cho21ms} summarized gamma-ray light curves of 12 novae 
(see their Figure 8).
Among them, six novae, V339 Del, V1369 Cen,
V5668 Sgr, V407 Lup, V5855 Sgr, and V5856 Sgr, 
clearly show that the first detection of $> 100$ MeV gamma-rays 
with $>2\sigma$ significance were almost just after optical maximum.

Also, in the classical nova YZ Ret, a close binary nova 
\citep[$P_{\rm orb}= 3.179$ hr;][]{schaefer22}, 
gamma-rays were detected probably after the optical peak  
\citep{sok22ll, kon22wa},
%\citep[e.g.,][for the gamma-ray light curve of YZ Ret]{sok22ll, kon22wa},
although the optical peak is not clearly identified.
% because of an observational gap around the $V$ maximum.
A few additional cases of gamma-ray novae in close binary includes 
V1324 Sco, of which gamma-ray detection is between the two optical maximum, 
and V357 Mus, of which optical maximum probably missed.  

In this way, the first detection dates of gamma-rays are 
consistent with the model prediction by \citet{hac22k}, 
that is, a shock arises just after the optical peak. 

%{\color{brown}
%The other novae may not support this conjecture.  However,
%the optical maximum of the nova V959 Mon was not observed,
%and the first gamma-ray detected dates were not identified for V392 Per
%and V906 Car because of Fermi/LAT solar panel failure.
%V407 Cyg is a symbiotic star, the companion of which is a Mira, not
%a close binary.  In V1324 Sco, the first gamma-ray detection is six days
%before the optical maximum, but there seem to be two optical peaks,
%the first peak is the reddest one in $B-V$ and about 8 days before
%the second 0.5 mag brighter peak in $V$ \citep[see Figure 1 of ][]{mun15wh}.
%If we adopt the first peak
%as the global optical maximum like in V339 Del, gamma-rays were first
%detected just after this optical maximum. 
%In V357 Mus, the first gamma-ray detection is four days before 
%optical maximum in their Figure 8 of \citet{cho21ms},
%but the exact optical maximum
%cannot be accurately determined because of 10 days $V$ data gap
%before their optical maximum \citep[see Figure 1 of ][]{mol20ib}.
%Thus, our conjecture on the starting date of a shock is broadly
%supported by a majority of gamma-ray detected novae.
%}

%V339 Del is also a close binary \citep[$P_{\rm orb}= 3.91$ hr;
%][]{schaefer22}.  GeV gamma-rays (filled green triangles) were first
%detected just after the global $V$ maximum, as shown in Figure 
%\ref{v339_del_v_skopal_wd_photo}b. 

\subsection{Dust shell formation}
\label{obs_dust_shell_formation}

A dust shell could be formed within a cool and dense shell
behind the radiative shock in a nova ejecta \citep[e.g.,][]{der17ml}.  
V339 Del shows rebrightening in the $K_{\rm s}$ light curve, 
suggesting a dust shell formation.  
Figure \ref{v339_del_v_bv_ub_color_curve_no2} shows that 
the $K_{\rm s}+3$ mag light curve has two peaks. After the first peak, 
it decreases along with the $V$ (or $y$) light curve. 
However, the $K_{\rm s}+3$ magnitude increases again 
from JD~2,456,552.5 (day $\sim 35$).
The $K_{\rm s}+3$ magnitude again peaked on JD~2,456,572.5 (day $\sim 55$). 
Also the $KLM$ band fluxes increase similarly 
\citep[see Figure 1 of][]{bur15a}.

This second increase is owing to dust formation. 
Dust formation is reported on UT 2013 September 27.8 
%%% 563.3 - 518.0 = 45.3
%%%\citep[JD~2456563.3,][]{shen13tt}, $\sim 45$ days after the outburst. 
by \citet{shen13tt}, $\sim 45$ days after the outburst. 
\citet{eva17bg} obtained the formation date of dust to be 34.75 days after
the eruption (their $t=0$ is UT 2013 August 13.9 $=$ JD~2,456,518.4).
The epoch of dust formation is indicated by a black arrow in Figure
\ref{v339_del_v_bv_ub_color_curve_no2}a.  

We suppose that a dust shell formed behind the shock
as shown in Figure \ref{wind_shock_config}d.  
The dust shell formation and its effects on other wavelength light curves 
will be discussed later in Section \ref{discussion}.

\subsection{SSS Phase}
\label{sss_phase}
The optically thick winds have stopped at the epoch of Figure 
\ref{wind_shock_config}c, because the mass of the hydrogen-rich envelope
decreases to a critical value due mainly to wind mass-loss. 
Then, the photospheric temperature becomes high enough to emit soft X-rays.
The nova entered the SSS phase \citep{kat94h, sal05h, wol13bb}.
Even if the winds stop emerging from the photosphere,
the latest winds are still colliding
with the shock and maintain the geometrically thin shocked shell,
as in Figure \ref{wind_shock_config}c.  

The SSS phase itself is not directly related to the shock, but
the optical depth of the shocked layer influences the penetration
of soft X-ray photons.
\citet{nel13} obtained the  absorption column density of
$N_{\rm H}\sim 1.8\times 10^{21}$~cm$^{-2}$ from the Chandra X-ray
spectrum, 89 days after the outburst, in the SSS phase.
This $N_{\rm H}$ is low enough for soft X-rays to penetrate the shocked
thin shell.  We will discuss the $N_{\rm H}$ value based on our shocked
shell model in Section \ref{section_hard_x-ray}.

\subsection{Nebular phase and SSS phase}
\label{nebular_sss_phases}

\citet{mun13dc} reported that the nova entered the nebular phase on 
%%%UT 2013 October 18 (JD~2,456,583.5: $\sim 65$ days after the outburst) 
UT 2013 October 18 ($\sim 65$ days after the outburst) 
from the flux ratio of [\ion{O}{3}]$/$H$\beta >1$.  
%%% 583.5 - 518.0 = 65.5
We indicate the epoch in Figure \ref{v339_del_v_bv_ub_color_curve_no2}. 
%\citet{sho16ms} reported that, 
%the ejecta were still in the transition to nebular on day 100, 
%during which time the Lyman series was still
%optically thick as with the electron density (Is this rewarding correct?).  

In the nebular phase, strong emission lines such as [\ion{O}{3}] 4957, 5007
\AA\  contribute to the $V$ band while the $y$ band does not include these
[\ion{O}{3}]  line region.
Thus, the $V$ and $y$ light curves depart from each other. 
Figure \ref{v339_del_v_bv_ub_color_curve_no2} shows such a deviation
between the $V$ and $y$ light curves.  These departing behaviors of $V$
and $y$ have been known in V1500 Cyg \citep{loc76m} and V1668 Cyg
\citep{gal80ko}. 

In V339 Del, the beginning of the nebular phase is almost coincident with 
the start of the SSS phase. 
\citet{mun15mm} discussed that the nebular emission could be 
excited by the SSS X-ray photons. 
They reported their Str\"omgren $b$, $y$, and narrow-band
H$\alpha$, [\ion{O}{3}] photometric evolution of V339 Del.  
They interpreted that,
as the ejecta becomes optically thin, 
% the spreading ionization caused by 
the supersoft X-ray input increases and 
the [\ion{O}{3}] 5007 \AA\  emission line rapidly grows in intensity. 
This causes the brightness in the [\ion{O}{3}] band to slow 
and then to stop the decline
(initial 20 days of the SSS phase), followed by an actual
and long lasting brightening \citep[see Figures 8 and 9 of ][]{mun15mm}.
At the end of the SSS phase (130 days past it begun), V339 Del
is still brighter by 0.2 mag in the [\ion{O}{3}] band than before it
started. 

We will discuss if the source of nebular emission is the WD X-ray emission 
or the shocked shell contribution based on our light curve model 
in Section \ref{discussion}.

\subsection{Coexistence of Dust and SSS Phases}
\label{coexist_dust_x-ray_raidation}

\citet{geh15eh} pointed out that V339 Del is a rare object 
among many novae because it shows both dust formation and 
a soft X-ray radiation in the same period.  
\citet{eva17bg} also argued the coexistence of X-ray radiation
and dust, and concluded that grain shattering by electrostatic stress
destroyed dust around V339 Del.
In other novae, the dust has not been exhibited
because a soft X-ray radiation is expected to have destroyed dust. 
We will elucidate the mechanism of this coexistence based on 
our shocked shell model in Section \ref{discussion}.

\section{Light Curve Model}
\label{light_curve_fitting}

Here we present a theoretical light curve model for the V339 Del outburst
assuming spherical symmetry.  We do not take into account
non-spherical nature of ejecta, mainly because we try to explain
the main properties of multiwavelength light curves. 
Non-sphericity of ejecta could be important to analyze complicated
line profiles in nova spectra, which is however beyond the scope
of the present work.

%{\bf (original) We assume spherical symmetry throughout this section.  
%As for multiwavelength
%light curve analysis, we do not feel that non-sphericity plays a crucial 
%role.  But, non-sphericity of ejecta could play an important role,
%as studied, in the analysis on emission/absorption line profiles
%\citep[e.g.,][for V339 Del]{sko14dt, sho16ms, kaw19sa}.}

Among a large set of nova model light curves presented by Hachisu and Kato, 
we chose a $M_{\rm WD}=1.25 ~M_\sun$ model 
with the chemical composition of the envelope ``neon nova 2 (Ne2),'' 
i.e., $(X, Y, Z, X_{\rm CNO},X_{\rm Ne})= 
(0.55, 0.30, 0.02, 0.10, 0.03)$.
Here, $X$, $Y$, and $Z$ are the mass fraction of hydrogen, helium, 
and heavy elements, $X_{\rm CNO}$ the total abundance of extra carbon, 
nitrogen, and oxygen, and $X_{\rm Ne}$ the extra neon 
\citep[e.g., ][]{hac06kb}. 
The optical light curve is tabulated in Table 3 of 
\citet{hac10k}.
%%  We describe our method in more details in Appendix \ref{theory}.

%Fig.5
%\placefigure{v339_del_v_logscale_no3}

%Fig.5  

\begin{figure}
%%\epsscale{0.75}
%%\epsscale{0.8}
%%\epsscale{1.0}
\epsscale{1.15}
\plotone{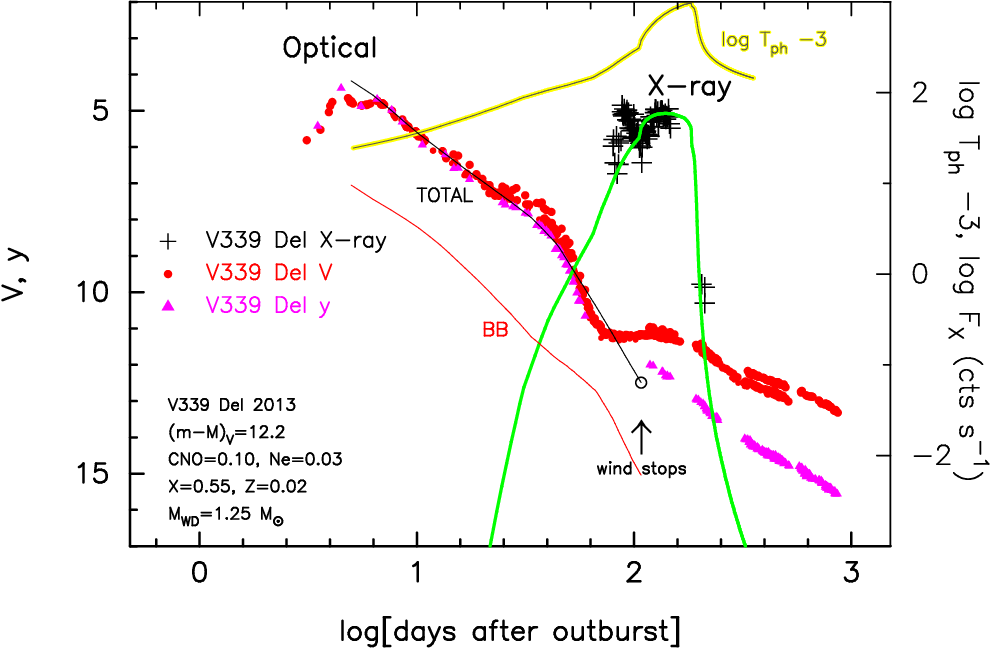}
%\plotone{v339_del_v_logscale_no3.epsi}
%\plotfiddle{evolution1.ps}{5.0cm}{270}{0.4}{0.4}{-170}{220}
\caption{
Our best-fit $1.25 ~M_\sun$ WD (Ne2) model light curves for $V$
and supersoft X-ray. 
The red line labeled BB denotes the blackbody flux $L_{V, \rm ph}$,
the black line labeled TOTAL the total $V$ flux of FF+BB,
calculated by Equation (\ref{luminosity_summation_flux_v-band}). 
The green line shows the soft X-ray flux (0.3--2.0 keV) for 
the blackbody assumption \citep{kat94h}.  
The yellow+gray line shows the photospheric temperature
($\log T_{\rm ph} ~({\rm K}) -3$).
Note that the black crosses
depict the X-ray count rates \citep[0.3--10.0 keV,][]{eva09}
which are not the same as the X-ray flux (green line). 
The observational data are the same as those
in Figure \ref{v339_del_v_bv_ub_color_curve_no2}.
\label{v339_del_v_logscale_no3}}
\end{figure}

\subsection{Optical $V$ Light Curve}
\label{v_light_curve_fitting}

\citet{hac06kb} proposed a formula of free-free flux in the $V$ band 
from nova winds based on \citet{kat94h}'s nova outburst model,
which can be simplified as 
\begin{equation}
L_{V, \rm ff,wind} = A_{\rm ff} ~{{\dot{M}^2_{\rm wind}}
\over{v^2_{\rm ph} R_{\rm ph}}},
\label{free-free_flux_v-band}
\end{equation}
where $\dot{M}_{\rm wind}$, $v_{\rm ph}$, and $R_{\rm ph}$ are
the wind mass-loss rate, wind velocity at the photosphere, and
photospheric radius, respectively, and $A_{\rm ff}$ is the 
coefficient for free-free emission.    
This luminosity represent the flux of free-free emission from optically thin
plasma just outside the photosphere.   See \citet{hac06kb} for
the derivation of this formula and \citet{hac20skhs} for details of
the coefficient $A_{\rm ff}$.   Note that optically-thick nova winds
become optically thin outside the photosphere 
(Figure \ref{wind_shock_config}a and b) and we obtain the free-free
flux using the physical values at the photosphere.

The total $V$-band flux is the summation of the free-free emission luminosity
and the $V$-band flux of the photospheric luminosity $L_{\rm ph}$,
\begin{equation}
L_{V, \rm total} = L_{V, \rm ff,wind} + L_{V, \rm ph}.
\label{luminosity_summation_flux_v-band}
% equation.2
\end{equation}
The photospheric $V$-band luminosity is calculated from a blackbody with
$T_{\rm ph}$ and $L_{\rm ph}$ using a canonical response function
of the $V$-band filter, where $T_{\rm ph}$ and $L_{\rm ph}$ are
the photospheric temperature and luminosity of our steady-state
envelope solution \citep[e.g., ][]{kat94h}.

To summarize, optically-thick winds are accelerated deep inside the
photosphere.  The wind becomes optically thin outside the photosphere
(Figure \ref{wind_shock_config}a and b).
The emission from the photosphere $L_{V, \rm ph}$
is calculated assuming blackbody with the photospheric temperature
$T_{\rm ph}$ and luminosity $L_{\rm ph}$.  On the other hand,
the wind continuously blows through the photosphere, the gas of which
is hot and optically thin outside the photosphere and emits photons
by free-free emission $L_{V, \rm ff, wind}$.  In other words,
$L_{V, \rm ff, wind}$ comes from plasma outside the photosphere.
Therefore, $L_{V, \rm ff, wind}$ is not limited by the Eddington
luminosity, which can operate only in the optically-thick region. 

Figure \ref{v339_del_v_logscale_no3} shows our model light curves of
the 1.25 $M_\sun$ WD (Ne2) fitted with the $V$ and supersoft X-ray
observation.  The blackbody flux from the photosphere (red line
labeled BB) is much smaller than that of free-free emission of ejecta, 
so the total flux (black line labeled TOTAL) is very close to 
the free-free emission luminosity. 
This tendency is also seen in other novae 
\citep[e.g., V1668 Cyg, V693 CrA, QU Vul, V351 Pup, V1974 Cyg, and 
V382 Vel in][]{hac16k}. 
It should be noted that, because the steady-state winds in \citet{kat94h}'s
model are valid in the decay phase of novae, our model light curve can be
applied to the post-maximum phase of novae.  Thus, we plot our model
light curve for the decay phase of a nova, i.e., after the optical maximum.

The TOTAL light curve well reproduce the observed data 
for both $V$ and $y$ until day $\sim 80$. 
The free-free light curve decreases with time but its decline 
rate changes on day $\sim 45$. 
This is because the wind mass loss rate quickly decreases 
with time when the photospheric temperature increases to 
$\log T$ (K) $\sim 5.2$ and the wind acceleration becomes 
ineffective (or inefficient).

%The brightness of a nova depends mainly on the free-free flux from winds.
The free-free flux of wind is determined mainly by the wind mass-loss rate
(Equation (\ref{free-free_flux_v-band})).
Therefore, the steep decrease in the brightness is due to the quick drop
in the wind mass-loss rate.  The radiative pressure gradients in the
envelope depend on the height of the iron peak
at $\log T ({\rm K})\sim 5.2$ of the OPAL opacity \citep{igl96r}
because the peak blocks radiative flux and makes
large radiative pressure gradients.
When the photospheric temperature is increasing to pass
through $\log T ({\rm K})\sim 5.2$ \citep[iron peak of the OPAL 
opacity, see $\log T$-$\kappa$ plot in Figure 2 of ][]{kat94h},
the effect of radiative pressure gradient is rapidly decreasing,
but is still working.
This corresponds to the steep decreasing phase in the brightness after day 45.
At about $\log T ~({\rm K}) \sim 5.5$ (see Figure
\ref{v339_del_v_logscale_no3}), the height of a residual part of the
iron peak inside the photosphere becomes too low (or small)
to drive optically-thick winds mainly because a large part of the iron peak
goes outside the photosphere.  Optically-thick winds driven by 
radiative pressure gradients stop at/around
$\log T ~({\rm K}) \sim 5.5$.
The date of wind stopping is denoted by the black arrow labeled
``wind stops.''

Thus, the quick decrease in the light curve during day $\sim 45$-80 
is owing to the quick decay in the wind mass-loss rate, 
not necessary owing to dust blackout. 
Such a quick decay is seen in the $y$ light curves of V1500 Cyg 
\citep{loc76m} and V1668 Cyg \citep{gal80ko}. 

Our $1.25 ~M_\sun$ WD model has a hydrogen-rich envelope of mass
$M_{\rm env}\approx 0.5\times 10^{-5} ~M_\sun$ at optical maximum.
If the mass-accretion rate to the WD is about 
$\dot{M}_{\rm acc}\sim 3\times 10^{-9} ~M_\sun$ yr$^{-1}$
(Figure \ref{vmax_t3_vmax_t2_selvelli2019_schaefer2018_2fig}),
the recurrence time is about $t_{\rm rec}\approx
(M_{\rm env}/\dot{M}_{\rm acc}) \sim 1,600$ yr.  These values are
broadly consistent with the positions at the MMRD diagrams 
(Figure \ref{vmax_t3_vmax_t2_selvelli2019_schaefer2018_2fig}).
%%%%Section \ref{mmrd_relation}. 
We assume that $(m-M)_V=12.2$ in Figure \ref{v339_del_v_logscale_no3}.
Then, our model light curve follows well the observed
$V$ and $y$ light curves.  This supports our adopted distance
of $d=2.1\pm 0.2$ kpc for $E(B-V)=0.18$, because 
$(m-M)_V= 5\log (d/{\rm 10~pc}) + 3.1 E(B-V)= 5\log 210 + 3.1\times 0.18=
12.17$.

\subsection{Supersoft X-Ray Light Curve}
\label{soft_x_light_curve_fitting}

The supersoft X-ray flux $L_{\rm X}$ is calculated from 
the photospheric temperature $T_{\rm ph}$ and luminosity $L_{\rm ph}$ 
of the same model as that of the optical light curve. 
The green line in Figure \ref{v339_del_v_logscale_no3} denotes 
the supersoft X-ray flux $L_{\rm X}$ for the 0.3--2.0 keV energy range. 
The X-ray flux rises as the photospheric temperature increases to 
$\log T$ (K) $> 5.2$ and keeps a constant value around the maximum 
and begins to decay after hydrogen burning turned off (day $\sim 180$). 

We fit our X-ray light curve to the Swift X-ray count rates (0.3--10.0 keV)
to set the flat peak on the top count rates, as shown 
in Figure \ref{v339_del_v_logscale_no3}.
The difference in the higher energy range, i.e., between 2.0 -- 10.0 keV, 
can be neglected, because most energies of X-rays are emitted at $< 0.7$ keV
\citep{osb13pb}.  Our model light curve is the X-ray luminosity $L_{\rm X}$, 
while the observational data are the count rate.  These two are not
the same, but our model X-ray light curve reasonably reproduces the 
temporal variation of the Swift/XRT count rates.

The X-ray count rate in the early SSS phase
shows a large amplitude variation (LAV) \citep{bea13op}.
This LAV disappears in the later SSS phase. 
\citet{sho16ms} reported a period of LAV from day 72 to 88, 
and then a decrease in the count rate by a factor of 5
from day 88 to day 108.
Figure \ref{v339_del_v_logscale_no3} shows 
that the count rate varies along our theoretical model.  
The LAV appears on day 72 and ends on 110, 
of which the period exactly corresponds to the phase 
when the wind mass-loss rate rapidly decreases to finally stop  
on day 108.  

Thus, we regard that the LAV is closely related to the condition of the 
optically-thick winds.
The timescale of the LAV is about 1 day \citep{bea13op}.  The thermal
timescale of the WD envelope is also about $\tau_{\rm th} \equiv
E_{\rm th}/L_{\rm ph} \sim (1\times 10^{43} {\rm ~erg} /
1\times 10^{38} {\rm erg~s}^{-1}) \sim 2$ days, where $\tau_{\rm th}$
and $E_{\rm th}$ are the thermal timescale and thermal energy of
the hydrogen-rich envelope of the WD, respectively.
Thus, the variation timescale is almost the same as the thermal
timescale of the envelope.  If the winds are thermally unstable just before
they stop, we are able to explain the timescale of the variability. 
See \citet{schw11}, \citet{osb11pb}, and \citet{mun15mm}
for other possible explanations on the various LAVs of novae.

Recently, \citet{mil23p} presented X-ray spectra observed with Chandra
on UT 2013 November 9 and 2013 December 6 (day 88 and 115) during
%%% JD 2456605.5 - 518.0= 87.5
%%% JD 2456632.5 - 518.0= 114.5 
the SSS phase (day 71 -- 200).
They obtained the atmospheric effective temperature of $T_{\rm eff}=
6.4\times 10^5$ K and a shell with the velocity of $-1400$ km s$^{-1}$
from absorption spectra.
This velocity is consistent with the diffuse enhanced system
of $-1400$ km s$^{-1}$ already discussed in Section 
\ref{post_maximum_phase}.

In our $1.25 ~M_\sun$ WD (Ne2) model, optically-thick wind is still blowing
until day $\sim 108$.  The velocity of the wind is regarded as that of the
diffuse enhanced system, $\sim -1400$ km s$^{-1}$.
Thus, the main absorption feature of the X-ray spectra is consistent with
our wind model.  The effective temperature of the WD
photosphere is about $T_{\rm ph}= 3\times 10^5$ K during day 88 to 115,
much lower than \citet{mil23p}'s value.    It should be noted that 
\citet{nel13} reported a temperature of k$T\sim 27$~eV
($T\sim 3.1\times 10^5$~K) for the Chandra X-ray spectrum of V339~Del 
on day 88, which is consistent with our $1.25 ~M_\sun$ WD (Ne2) model. 
%%%= JD 2,456,606.25
%%%$\sim 89$ days after the outburst.

On day 88, the photospheric radius is about 16 times larger than that
of the original (cool) WD.
The difference by a factor of two in the temperature comes from 
the fact that \citet{mil23p} adopted \citet{rau03}'s atmosphere model,
which is based on the assumptions of plane-parallel and in hydrostatic
balance.
%, and therefore,
%his atmospheric model cannot be applied to this kind of
%extended atmosphere with winds.

\citet{mil23p} further claimed a second shell of mass
$\sim 5\times 10^{-4} ~M_\sun$ with velocity of $-4000$ km s$^{-1}$.
This second shell is very unlikely because no such high velocity
components were observed in optical spectroscopy of V339 Del
\citep[e.g.,][]{sko14dt, deg15sm, sho16ms}.  Also,
their obtained shell mass of $\sim 5\times 10^{-4} ~M_\sun$
is too massive to be compatible with our ejecta mass of $M_{\rm shell}= 0.45
\times 10^{-5} ~M_\sun$, as will be discussed in Section
\ref{section_hard_x-ray}.
We may conclude that this second shell is spurious.

%xxxxxxxxxxxxxxxxxxxxxxxxxxxxxxx 

%Fig.6  a,b,c (three figures vertical)
%\placefigure{density_pp_interaction_v339_del}

\begin{figure}
\gridline{\fig{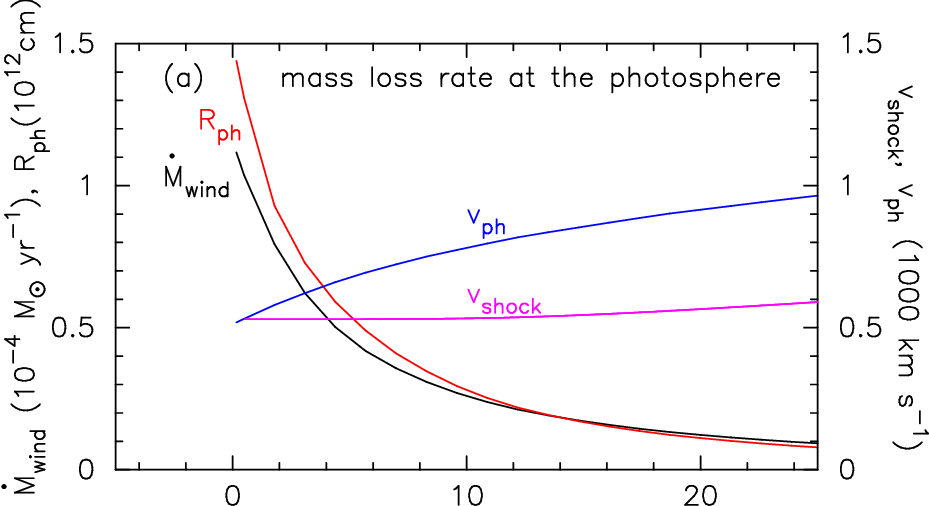}{0.47\textwidth}{}
%%%          \fig{f1.eps}{1.0\textwidth}{(a)}
          }
\gridline{
%%%\fig{f1.eps}{1.0\textwidth}{(b)}
          \fig{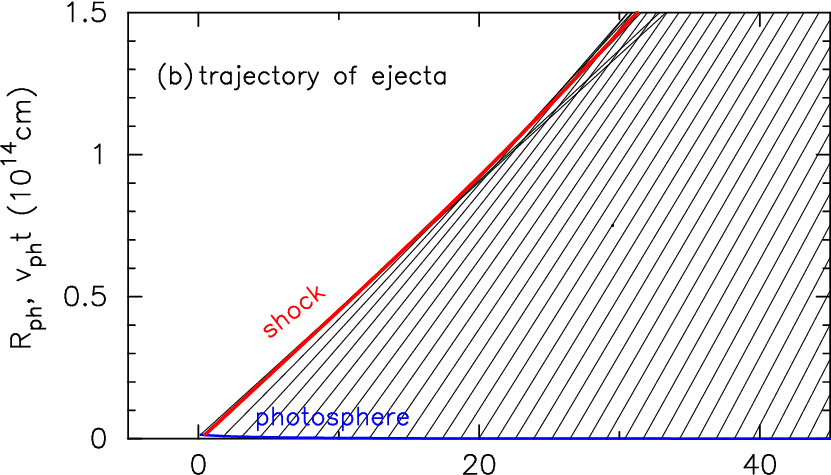}{0.42\textwidth}{}
          }
\gridline{
%%%\fig{f1.eps}{1.0\textwidth}{(c)}
          \fig{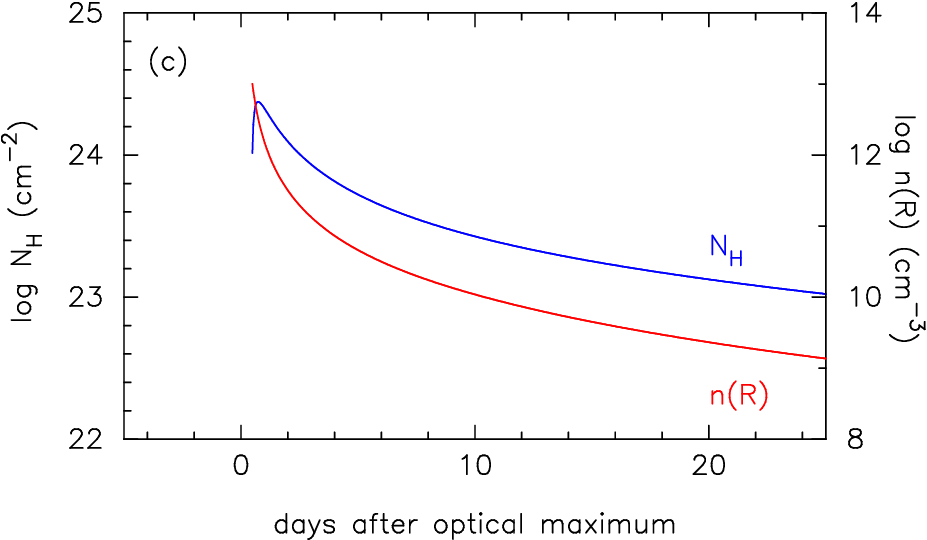}{0.47\textwidth}{}
          }
%\begin{figure}
%\epsscale{1.15}
%%\rotate
%%\plotone{f317.eps}
%\plotone{f6.eps}
%\plotone{../../v1668_cyg/steadystate/mdot_radius_velocity_v339_del.epsi}
%\plotone{../../v1668_cyg/steadystate/trajectory_wind_v339_del.epsi}
%\plotone{../../v1668_cyg/steadystate/density_pp_interaction_v339_del_no2.epsi}
%\plotfiddle{evolution1.ps}{5.0cm}{270}{0.4}{0.4}{-170}{220}
\caption{
Shock properties for our 1.25 $M_\sun$ WD (Ne2) model.
Only the decay phase (after the optical maximum) is plotted;
the time $t=0$ correspond to the optical maximum.
(a) Various photospheric properties:
the wind mass-loss rate (black line,
labeled ${\dot M}_{\rm wind}$), wind velocity (blue, $v_{\rm ph}$),
photospheric radius (red, $R_{\rm ph}$), shock velocity
(magenta, $v_{\rm shock}$).
(b) Trajectories (straight black lines) of winds ejected at each
time from the photosphere.  A strong shock (solid red line)
arises soon after the optical maximum and travels outward
at a speed of $\sim 500$ -- 600 km s$^{-1}$.
The blue line shows the position of the photosphere.
(c) Temporal variation in the hydrogen column density $N_{\rm H}$
(blue line) behind the shock, and the number density $n(R)$
(red line) just in front of the shock ($R=R_{\rm sh}$).
\label{density_pp_interaction_v339_del}}
\end{figure}

%\subsection{Multiple velocity systems in the ejecta}
%\label{multiple_velocities_ejecta}

\subsection{A shock formation far outside the photosphere}
\label{shock_formation_outside_photosphere}

Figure \ref{density_pp_interaction_v339_del}a shows 
the temporal variation of the wind velocity at the photosphere
$v_{\rm ph}$ of our $1.25 ~M_\sun$ WD (Ne2) model. 
The velocity increases with time, so the matter ejected later
will catch up the matter ejected earlier and this results in formation
of a strong shock. 

Figure \ref{density_pp_interaction_v339_del}b depicts the trajectories
of each wind particle. Here we assume a ballistic motion 
for each fluid particle after it is ejected from the photosphere. 
The position of each fluid particle is denoted by the black line. 
The shock wave expands far outside the photosphere. 
We use mass and momentum conservations to derive the shock velocity.
The method of this shock calculation is briefly described in \citet{hac22k}.

Figure \ref{density_pp_interaction_v339_del}c shows the column density of
hydrogen $N_{\rm H}$ behind the shock and the number density of plasma 
particles $n(R)$ in just front of the shock. We will use these values 
to estimate the X-ray and gamma-ray emission from the shocked matter.  

\citet{hac22k} showed that the shock properties and high energy emissions 
in a classical nova can be well explained if we adopt observed velocities
instead of their theoretical values. 
We have already obtained the pre-maximum, principal, and diffuse-enhanced
absorption line systems in Section \ref{post_maximum_phase}.  These
line systems can be interpreted to originate from the earliest wind,
shocked shell, and inner wind, respectively, as in Figure
\ref{wind_shock_config}b.

In what follows, we regard the principal and diffuse-enhanced
velocity systems to be $v_{\rm p}= 800$ km s$^{-1}$ and
$v_{\rm d}= 1400$ km s$^{-1}$, respectively, instead of
our shock model velocities of $v_{\rm shock}= 500$--600 km s$^{-1}$ and
$v_{\rm wind}= 500$--1000 km s$^{-1}$.
Here, $v_{\rm p}$ is the velocity of the principal system, and
$v_{\rm d}$ is that of the diffuse-enhanced system.

Our $1.25 ~M_\sun$ WD (Ne2) model shows the shock velocity of
$v_{\rm shock}\sim 500$--600 km s$^{-1}$ and the wind velocity of
$v_{\rm ph}\sim 500$--1000 km s$^{-1}$, as in Figure 
\ref{density_pp_interaction_v339_del}a.  These velocities are
about 40\% smaller than the observed ones.  The wind acceleration
depends on the opacity in our optically-thick wind theory.
When the radiative opacity was changed to the OPAL opacity \citep{igl96r},
the wind velocity increases drastically.  Thus, we feel a slight lack of
acceleration agent to obtain the wind velocities comparative with the 
observation \citep[see, e.g.,][for a possible cause 
for higher opacity]{bai14nl}.

\subsection{Duration of shock}
\label{shock_duration_hard_x-ray}

The shock arises just after the optical peak 
(see Figure \ref{density_pp_interaction_v339_del}).  
This shock continues until shortly after the wind stops, 
and disappears when the last wind reaches the shock front, as
illustrated in Figure \ref{wind_shock_config}c.
The lifetime of the shock can be estimated by the same way as
that of \citet{hac22k, hac23k}:
\begin{equation}
\tau_{\rm shock}= {{t_{\rm ws}} \over
{\left( 1- {{v_{\rm p}} \over {v_{\rm d}}}\right)}}.
\label{duration_of_shock}
\end{equation}
Substituting $v_{\rm shock}\approx v_{\rm p}=800$ km s$^{-1}$
(principal system), $v_{\rm ph}\approx v_{\rm d}=1400$ km s$^{-1}$
(diffuse-enhanced system), and $t_{\rm ws}=103$ days ($=108 - 5$ days;
the epoch when the optically-thick winds stop in our model minus the date
of the global optical maximum) into Equation (\ref{duration_of_shock}),
we obtain the shock duration is $\tau_{\rm shock}= 103/0.429= 240$ days.
Therefore, we expect hard X-ray emission until day $\sim 245$.
%%% 245 days -- log 245 = 2.39
This date is close to the end of the SSS phase on day $225$.  

In some novae like in YZ Ret 2020 \citep[e.g., Figure 1 of ][]{hac23k}
and V2491 Cyg 2008 \citep[e.g., Figure 7 of ][]{kat21sh},
the X-ray count rate does not drop to zero even after the hydrogen
shell-burning ended.  These late X-rays could originate from the shock
(or shocked shell).  In V339 Del, the X-ray count rate rapidly decreased
to undetected level at the end of the SSS phase, which is consistent with
the date when the shock disappeared.

\subsection{Hard X-ray emission from shocked matter}
\label{section_hard_x-ray}

The temperature just behind the shock is estimated to be
\begin{eqnarray}
kT_{\rm shock}& \sim & {3 \over 16} \mu m_p
\left( v_{\rm wind} - v_{\rm shock} \right)^2 \cr
& \approx & 1.0 {\rm ~keV~}
\left( {{v_{\rm wind} - v_{\rm shock}} \over
{1000 {\rm ~km~s}^{-1}}} \right)^2,
\label{shock_kev_energy}
\end{eqnarray}
where $k$ is the Boltzmann constant,
$T_{\rm shock}$ is the temperature just after the shock
\citep[see, e.g.,][]{met14hv},
$\mu$ is the mean molecular weight ($\mu =0.5$ for hydrogen plasma),
and $m_p$ is the proton mass.
Substituting $v_{\rm shock}= v_{\rm p}=800$ km s$^{-1}$ and
$v_{\rm wind}= v_{\rm d}=1400$ km s$^{-1}$,
we obtain the post-shock temperature
$k T_{\rm shock}\sim 0.36$ keV.

Mechanical energy of the wind is converted to thermal energy
by the reverse shock \citep{met14hv} as
\begin{eqnarray}
L_{\rm sh}& \sim & {{9}\over {32}} {\dot M}_{\rm wind}
{{( v_{\rm wind} - v_{\rm shock} )^3} \over {v_{\rm wind}}} \cr
&=& 1.8\times 10^{37}{\rm ~erg~s}^{-1}
\left( {{{\dot M}_{\rm wind}} \over
{10^{-4} ~M_\sun {\rm ~yr}^{-1}}} \right) \cr
 &  & \times
\left( {{{v_{\rm wind} - v_{\rm shock}} \over {1000{\rm ~km~s}^{-1}}}}
\right)^3
\left( {{{1000{\rm ~km~s}^{-1}} \over {v_{\rm wind}}} }\right).
\label{shocked_energy_generation}
\end{eqnarray}
Substituting $\dot{M}_{\rm wind}= 1.4 \times 10^{-4} ~M_\sun$ yr$^{-1}$
from Figure \ref{density_pp_interaction_v339_del}a,
we obtain the post-shock energy of
$L_{\rm sh} \sim 6.8\times 10^{36}$ erg s$^{-1}$.

In the later nebular and SSS phases, the velocity and mass of the
shocked shell do not change so much.
The column density of hydrogen is estimated from
$M_{\rm shell}= 4 \pi R_{\rm shell}^2 \rho h_{\rm shell}$,
where $\rho$ is the density in the shocked shell,
and $h_{\rm shell}$ is the thickness of the shocked shell.
If we take an averaged velocity of shell $v_{\rm shock}=
v_{\rm shell}$, the shock radius is calculated from 
$R_{\rm shell}(t)= v_{\rm shock}\times t$.  This reads
\begin{eqnarray}
N_{\rm H} & = & {{X \over m_p} {{ M_{\rm shell} }
\over {4 \pi R^2_{\rm shell}}}} \cr
 & \approx & 4.8\times 10^{22} {\rm ~cm}^{-2}
\left({X \over {0.5}}\right)
\left( {{M_{\rm shell}} \over {10^{-5} M_\sun}} \right)
\left( {{R_{\rm shell}} \over {10^{14} {\rm ~cm}}} \right)^{-2}
\cr
 & \approx & 6.4 \times 10^{20} {\rm ~cm}^{-2}
\left({X \over {0.5}}\right)
\left( {{M_{\rm shell}} \over {10^{-5} M_\sun}} \right) \cr
& & \times
\left( {{v_{\rm shell}} \over {1000 {\rm ~km~s}^{-1}}} \right)^{-2}
\left( {{t} \over {100~{\rm days}}} \right)^{-2}.
\label{column_density_hydrogen_time}
\end{eqnarray}
This gives $N_{\rm H}\approx 8\times 10^{20}$ cm$^{-2}$ for
$M_{\rm shell}= 0.45\times 10^{-5} ~M_\sun$,
$v_{\rm shell}=800$ km s$^{-1}$, and $t=80$ days, which is close to
the start day of the SSS phase.  This $N_{\rm H}$ value is broadly
consistent with $N_{\rm H}\sim 1.8\times 10^{21}$~cm$^{-2}$ obtained
by \citet{nel13} from the Chandra X-ray spectrum on day 88.

For hard X-ray observations,
\citet{pag13b} reported the Swift/XRT detection of X-rays mainly below 2 keV
energies between day 37 and day 40. They estimated the hydrogen column
density of $N_{\rm H}=4.9^{+7.5}_{-3.2} \times 10^{22}$ cm$^{-2}$.
Our 1.25 $M_\sun$ WD (Ne2) model gives a similar column density of 
$N_{\rm H}\sim 1.1\times 10^{23}$ cm$^{-2}$ on day 40.

%Fig.7
%\placefigure{v339_del_v_gamma_logscale_no2.epsi}

\begin{figure}
\epsscale{1.15}
%%\rotate
\plotone{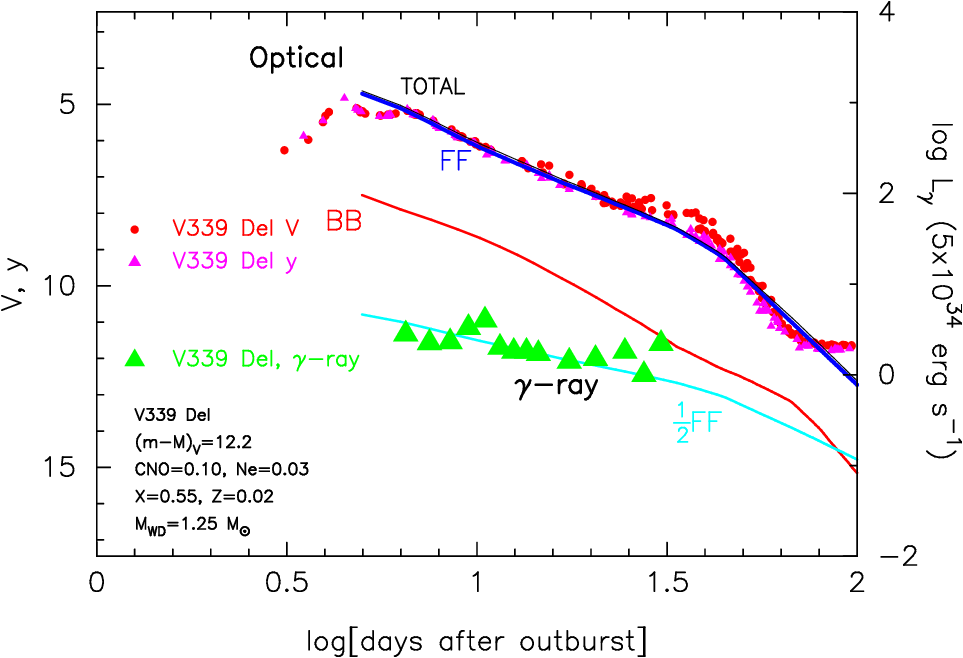}
%\plotone{v339_del_v_gamma_logscale_no2.epsi}
%\plotfiddle{evolution1.ps}{5.0cm}{270}{0.4}{0.4}{-170}{220}
\caption{
The optical $V$ (filled red circles) and $y$ (filled magenta triangles)
as well as the GeV gamma-ray flux (large filled green triangles)
light curves of V339 Del 2013. The distance modulus in the $V$ band of
$\mu_V\equiv (m-M)_V= 12.2$ is assumed.
We also plot our model light curves of the 1.25 $M_\sun$ WD (Ne2):
The thick blue line labeled FF denotes the free-free flux, 
$L_{V,{\rm ff, wind}}$, calculated from Equation 
(\ref{free-free_flux_v-band}), the red line labeled BB 
corresponds to the photospheric brightness, $L_{V,{\rm ph}}$,
with blackbody assumption, and the black line labeled TOTAL is
the sum of FF$+$BB.  We add the ${ 1 \over 2} m_{\rm ff}$ model
light curve (cyan line labeled ${ 1 \over 2}$FF, calculated from equation
(\ref{gamma_ray_half_magnitude})) with an arbitrary shift in the
vertical direction to match the decay trend of GeV gamma-ray.
\label{v339_del_v_gamma_logscale_no2}}
\end{figure}

\subsection{GeV gamma-ray emission from shocked matter}
\label{gev_gamma_ray_emission}

GeV gamma-rays were observed in the V339 Del outburst with 
the Fermi/LAT \citep{ack14aa}.  Such GeV gamma-rays originate
from strong shocks \citep[see, e.g.,][for a recent review]{cho21ms}.
The positive detection of gamma-rays started on $t\sim 5$ days, as shown
in Figure \ref{v339_del_v_skopal_wd_photo}b,
just after the global $V$ maximum.  This epoch of appearance is
consistent with our shock model \citep{hac22k}.

\citet{hac22k} obtained the decay trend of gamma-ray flux based on 
their shock model.  In their model, the optical flux is dominated
by free-free emission and is given by $L_{V, \rm ff,wind}\propto 
(\dot{M}_{\rm wind}/v_{\rm ph})^2 \propto f(t)$ (Equation 
(\ref{free-free_flux_v-band})) while the gamma-ray flux is given by
$L_{\gamma, \rm sh} \propto (\dot{M}_{\rm wind}/v_{\rm wind})\propto
[f(t)]^{1/2}$ (Equation (\ref{shocked_energy_generation})),
where $f(t)$ is a function of time $t$.
Therefore, the light curves (in magnitude) are written as
\begin{eqnarray}
m_{\gamma, \rm sh}(t) &=& -2.5 \log L_{\gamma, \rm sh} + {\rm ~const.} \cr
&=& -2.5 \log [f(t)]^{1/2} + {\rm ~const.} \cr
&=& {1\over 2}\left(-2.5 \log f(t) + {\rm ~const.}\right) \cr
&=& {1\over 2}(-2.5 \log L_{V, \rm ff,wind} + {\rm ~const.}) \cr
&=& {1\over 2} m_{V,\rm ff, wind}(t) + {\rm ~const}.,
\label{gamma_ray_half_magnitude}
\end{eqnarray}
so the decay slope is slow by a factor of 2.
%Here, the proportionality constant is included in the term ``const.''
We plot the $m_{\gamma, \rm sh}$ (cyan line labeled ${1 \over 2}$FF)
in Figure \ref{v339_del_v_gamma_logscale_no2}.  This model slope
reasonably represents the decay trend of GeV gamma-ray fluxes.
Similar decay trends were observed in V5855 Sgr \citep{hac22k}
and YZ Ret \citep{hac23k}. 

\citet{ack14aa} obtained the flux of GeV gamma-ray (0.1--300 GeV) to be
$L_{\gamma}= 2.6\times 10^{35}$ erg s$^{-1} (d/{\rm 4.2 ~kpc})^2$.
The optical luminosity is about $L_{\rm opt}= 3\times 10^{38}$ erg s$^{-1}
(d/{\rm 3~kpc})^2$ on the same day \citep{sko14dt}.
The gamma-to-optical ratio is $L_{\gamma}/L_{\rm opt} = 4\times 10^{-4}$,
if we adopt $d=2.1$ kpc \citep{bai21rf}.
The $L_{\gamma}/L_{\rm opt}$ of V339 Del is lowest among the gamma-ray 
detected novae \citep[typically $L_{\gamma}/L_{\rm opt}\sim 
0.001-0.01$;][]{li17mc, cho21ms}.  This could be due to a smaller difference
between $v_{\rm d}$ and $v_{\rm p}$, i.e., $v_{\rm d} - v_{\rm p}=600$
km s$^{-1}$, which results in a relatively weak shock of
$k T_{\rm shock}= 0.36$ keV in Equation (\ref{shock_kev_energy}).
On the other hand, other several novae showed $v_{\rm d} - v_{\rm p}
\gtrsim 1000$ km s$^{-1}$, which results in a relatively strong shock of
$k T_{\rm shock}\gtrsim 1$-2 keV \citep[e.g.,][]{hac22k}. 

The gamma-ray luminosity obtained by \citet{ack14aa} is $L_{\gamma}=
6.5\times 10^{34}$ erg s$^{-1}$ for $d=2.1$ kpc.  In our model, the shock
energy generation is obtained to be $L_{\rm sh}= 6.8\times 10^{36}$
erg s$^{-1}$ (Section \ref{section_hard_x-ray}).  The ratio of
$L_{\gamma}/L_{\rm sh}= 6.5\times 10^{34}/6.8\times 10^{36}
\approx 0.01$, about 1\% conversion rate from the shock energy to
the gamma-ray energy, is consistent with the conversion limit of
\begin{equation}
L_{\gamma}= \epsilon_{\rm nth} \epsilon_{\gamma} L_{\rm sh}
\lesssim 0.03 ~L_{\rm sh},
\end{equation}
where $\epsilon_{\rm nth}\lesssim 0.1$ is the fraction of the shocked
thermal energy to accelerate nonthermal particles, and
$\epsilon_{\gamma}\lesssim 0.1$ is the fraction
of this energy radiated in the Fermi/LAT band \citep[typically
$\epsilon_{\rm nth} \epsilon_{\gamma} < 0.03$;][]{met15fv}.

%Fig.8 ab (2x1 horizontal x vertical)
%\placefigure{optical_m10selfcon_abs_magnitudes}

\begin{figure*}
\gridline{\fig{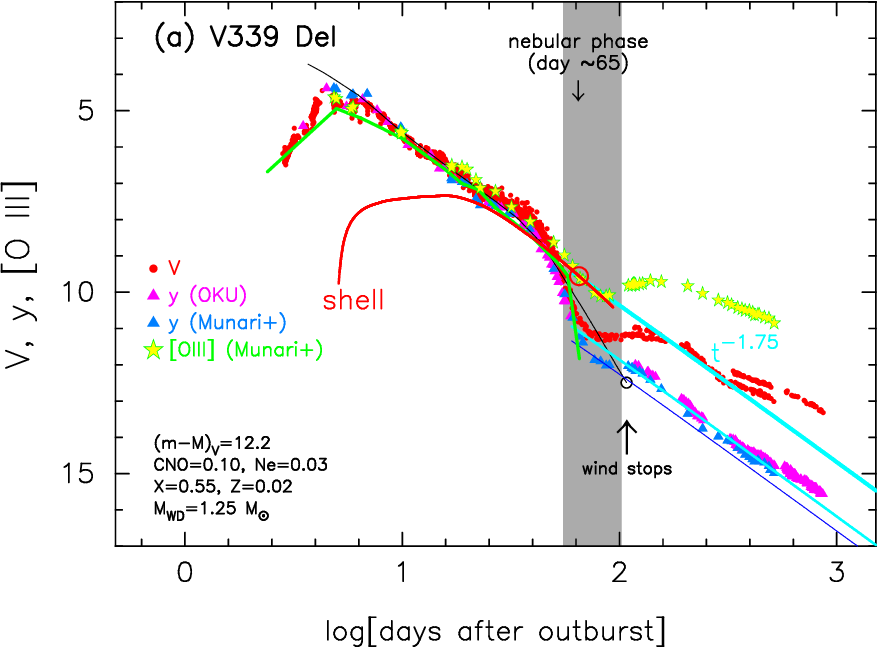}{0.5\textwidth}{}
          \fig{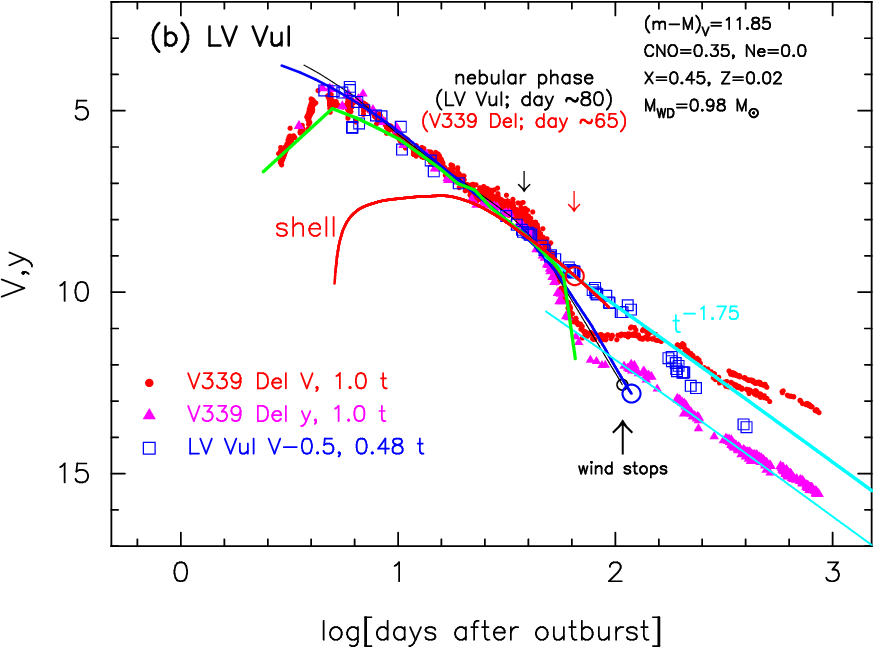}{0.5\textwidth}{}
          }
%\gridline{
%          \fig{f7c.eps}{0.5\textwidth}{}
%          \fig{f7d.eps}{0.5\textwidth}{}
%          }
%\begin{figure*}
%\epsscale{1.15}
%\epsscale{0.5}
%\plotone{f1016b.eps}
%\plotone{f1016b.eps}
%%\rotate
%\plotone{f4.eps}
%\plotone{optical_m10selfcon_abs_magnitudes.epsi}
%\plotone{v339_del_lv_vul_m10_shell_logscale_no3.epsi}
%\plotfiddle{evolution1.ps}{5.0cm}{270}{0.4}{0.4}{-170}{220}
\caption{
(a) The $V$, $y$, and [\ion{O}{3}] light curves of V339 Del.
In addition to the data of $V$ and $y$ (OKU) in Figure
\ref{v339_del_v_bv_ub_color_curve_no2}, 
we add the $V$ data (AAVSO), $y$ (Munari+) and [\ion{O}{3}] \citep{mun15mm}.
The gray-shaded region represents the substantial absorption phase
by a dust shell.  The black line is our 1.25 $M_\sun$ WD (Ne2) model
for the total FF+BB flux.  The optically thick wind ends at the small open
circle (on day 108).  The thick green line denotes the total FF+BB flux of
the fully self-consistent 1.0 $M_\sun$ WD model \citep{kat22sha, hac23k},
which is time-stretched by $\Delta\log t= -0.72$ and vertical shift of
$\Delta V= +0.5$ (see Section
\ref{v339_del_time_stretch} for the time-stretching method).
The red line labeled shell is the emission from the shocked shell
calculated from Equation (\ref{luminosity_shocked_shell_ff_flux}).
We add three lines of the universal decline law of $t^{-1.75}$
($V$: thick solid cyan line,
$y$: thin solid cyan line, $y$: thin solid blue line by dust absorption).
(b) Comparison between V339 Del and LV Vul with the time-stretching factor
of $\Delta\log t= -0.32$ and the $V$ shift of $\Delta V= -0.5$.
The $V$ data of LV Vul are the same as those in \citet{hac16k}.
The thick blue line denotes our FF+BB 
model light curve of a $0.98 ~M_\sun$ WD \citep[CO3;][]{hac16k} with
the same time-stretching as that for LV Vul. 
%(c) Same as in panel (b), but for V1668 Cyg.
%The data of V1668 Cyg ($V$ and $y$) are the same as those in 
%Figure 3 of \citet{hac16k}.  We add the visual magnitudes (small magenta-blue
%dots) of V1668 Cyg between day 90 and day 340 because $V$ data are too 
%few in this period.  The green line denotes our FF+BB
%model light curve of a $0.98 ~M_\sun$ WD \citep[CO3;][]{hac16k}.
%(d) Same as in panel (b), but for V1500 Cyg \citep{hac06kb, hac16k}.
%The green line denotes our FF+BB
%model light curve of a $1.2 ~M_\sun$ WD \citep[Ne2;][]{hac10k}.  
\label{optical_m10_abs_lv_vul_mag}}
\end{figure*}

\subsection{Emission from the Shocked Shell}
\label{emission_shocked_shell}

Figure \ref{v339_del_v_gamma_logscale_no2} shows that 
our theoretical $V$ light curve calculated by Equation 
(\ref{luminosity_summation_flux_v-band}) explains the decay phase of
the $V$ and $y$ light curves of V339 Del until day $\sim 60$. 
This means that the $V$ band flux is dominated by free-free emission
(blue line labeled FF) from the optically thin ejecta just outside 
the photosphere.  Note that the optically-thick winds become 
optically thin outside the photosphere (Figure
\ref{wind_shock_config}a and b).
The contribution from the photospheric emission 
(red line labeled BB) is rather small. 

In the later phase, the shocked shell far outside the photosphere
begins to contribute to the $V$ band flux.  
\citet{hac23k} estimated the luminosity of free-free emission 
from the shocked shell as
\begin{equation}
L_{V, \rm shell} = B_{\rm ff}
{{M_{\rm shell}^2} \over {4\pi R_{\rm shock}^2 h}},
\label{luminosity_shocked_shell_ff_flux}
\end{equation}
where $M_{\rm shell}$, $R_{\rm shock}$, and $h$ are the mass, radius,
and thickness of the shocked shell, respectively, and 
$B_{\rm ff}$ is the proportionality constant. 
We assume $h=$constant in time when the shock is alive.  
Note that this expression roughly includes the fluxes of bound-free
and bound-bound emission by replacing $B_{\rm ff}$ with $B_{\rm bf}$ or
$B_{\rm bb}$, where $B_{\rm bf}$ and $B_{\rm bb}$ are the coefficients
of bound-free and bound-bound emissions, respectively.  The other details
and derivation of Equation (\ref{luminosity_shocked_shell_ff_flux})
are described in Appendix B of \citet{hac23k}.

Now we have two different luminosities of
$L_{V, \rm total}= L_{V, \rm ff,wind} + L_{V, \rm ph}$
(FF+BB: equation (\ref{luminosity_summation_flux_v-band})) and
$L_{V, \rm shell}$ (shell: equation (\ref{luminosity_shocked_shell_ff_flux})).
\citet{hac23k} adopted 
\begin{equation}
L_V = \max\left( L_{V, \rm ff,wind}+L_{V, \rm ph},
~L_{V, \rm shell}\right),
\label{final_luminosity_wind_shell}
\end{equation}
for the total flux, instead of simple summation of the two,
because the shocked shell may absorb a part
of $L_{V, \rm ff,wind}+L_{V, \rm ph}$ and re-emit it
as a part of $L_{V, \rm shell}$.

We show the shell emission calculated from Equation 
(\ref{luminosity_shocked_shell_ff_flux}) in Figure 
\ref{optical_m10_abs_lv_vul_mag}a (thick red line labeled shell),
which is taken from \citet{hac23k}. 
The shocked shell emission gradually increases after the shock arises,
and reaches a plateau peak on day 16, 
followed by a slow decay approximately of
$L_V \propto t^{-1.75}$ (thick cyan line labeled $t^{-1.75}$).

\subsection{The Universal Decline Law}
\label{universal_decline_law}

\citet{hac06kb} found that theoretical light curves of free-free emission 
calculated based on the nova winds show 
a decline trend of $L_V \propto t^{-1.75}$, which is independent of the 
WD mass and chemical composition of the envelope. 
They and their following works showed that many novae show 
the $L_V \propto t^{-1.75}$ law. 
Because the photospheric blackbody flux is much smaller than the flux of
free-free emission, the summation of FF+BB fluxes also follows 
the universal decline law \citep{hac15k}. 

\citet{hac23k} further showed that the universal decline law of $t^{-1.75}$ 
can be extended over the nebular phase in which 
the optically thick wind has stopped, 
if the emission from the shocked shell is taken into account. 
They extended the universal decline law of $L_V \propto t^{-1.75}$
to the nebular phase until the shock disappears. 

Figure \ref{optical_m10_abs_lv_vul_mag}a shows the broad band $V$, 
intermediate band $y$, and narrow band [\ion{O}{3}] light curves
of V339 Del.  Both the $V$ and $y$ light curves similarly decline 
along with the universal decline law of $\propto t^{-1.75}$ 
until day $\sim 50$. After the quick decay in the dust blackout phase
(gray-shaded region), the $V$ light curve recovered on day $\sim 140$
and came back to the original decline trend of $\propto t^{-1.75}$
(thick cyan line labeled $t^{-1.75}$).  The $y$ band magnitude does not
recover.  After the dust blackout ended
(day $\sim 100$), it shows the same slope of
$L_y \propto t^{-1.75}$ (thin cyan line) starting from the bottom of the
dust dip.  The location of the $t^{-1.75}$ line (thin cyan line)
is about 1.5 mag below the original decline trend (thick cyan line).
The recovery of the $V$ light curve starts with the significant increase
of the [\ion{O}{3}] line emission, because the $L_V$ luminosity includes
the $L_{\rm [O\,III]}$ flux, but the $L_y$ luminosity does not.

Figure \ref{optical_m10_abs_lv_vul_mag}b shows the $V$ light curve of 
LV Vul overplotted with the $V$ and $y$ light curves of V339 Del. 
LV Vul evolves slower than V339 Del, so we applied 
the time-stretching method (see Section \ref{v339_del_time_stretch}) 
to overlap the same evolution stages of LV Vul with those of V339 Del.
The time-stretching factor is $f_{\rm s}=0.48$, that is, the time-shift by
$\Delta \log t = -0.32$ in the horizontal direction, and the vertical
shift by $\Delta V = -0.5$ mag, as shown by the inserted text
of ``LV Vul V-0.5, 0.48 t.'' 
LV Vul shows no dust formation. The light curve (blue square) declines almost 
along with the same universal decline law of $\propto t^{-1.75}$ 
(thick cyan line) over the nebular phase. 
It should be noted that the shell emission (red line) dominates
the $V$ luminosity in the nebular phase.
Comparing V339 Del (Figure \ref{optical_m10_abs_lv_vul_mag}a) with
LV Vul (Figure \ref{optical_m10_abs_lv_vul_mag}b), we understand
that the shell emission (red line) is dominated by [\ion{O}{3}] lines.
Thus, we regard that the coefficient in Equation 
(\ref{luminosity_shocked_shell_ff_flux}) is not $B_{\rm ff}$ (free-free)
but $B_{\rm bb}$ (bound-bound) for [\ion{O}{3}] lines in the nebular phase.

%Figure \ref{optical_m10_abs_lv_vul_mag}c compares the light curve of
%V1668 Cyg with those of V339 Del.  The evolution timescale of V1668 Cyg
%is very similar to that of LV Vul, so we adopt the same time-stretching
%factor $f_{\rm s}=0.48$.
%The $V$ magnitude declines along with the same $\propto t^{-1.75}$ line 
%as that of V339 Del.  Although there are few $V$ data of V1668 Cyg
%during the dust blackout period (gray-shaded region),  
%we can see the $y$ magnitude declines almost along with the model FF+BB
%light curve of a $0.98 ~M_\sun$ WD (CO3).

%Figure \ref{optical_m10_abs_lv_vul_mag}d shows V1500 Cyg against 
%V339 Del. The time-stretching factor is $f_{\rm s}=0.91$.
%The $V$ magnitude declines almost along with the $\propto t^{-1.75}$ line
%except for the very early phase until day $\sim 5$.  In the nebular phase,
%however, the $V$ data deviate from one another among several groups. 
%These splittings could be due to slight differences in their $V$ filter
%response functions \citep[see, e.g., Figure 1 of ][]{mun13dcvf}.
%The $y$ light curve decays broadly along with our FF+BB model light curve
%(blue line) calculated from a $1.2 ~M_\sun$ WD (Ne2) model \citep{hac10k}. 

To summarize, there are two main optical sources for $V$, one is the flux of
free-free (FF) emission from winds ($L_{V, \rm ff, wind}$
in Equation (\ref{free-free_flux_v-band})) that is dominant
in the early decay phase, and the other is the flux of free-free, bound-free,
and bound-bound emissions from shocked shell ($L_{V, \rm shell}$ in 
Equation (\ref{luminosity_shocked_shell_ff_flux}))
that is dominant in the nebular phase or later.
\citet{hac23k} found that the shocked-shell flux of $L_{V, \rm shell}$
exceeds the free-free flux of $L_{V, \rm ff, wind}$ in the nebular phase.
These two fluxes are connected smoothly to the universal decline law
of $\max(L_{V, \rm ff, wind}, L_{V, \rm shell})= L_V \propto t^{-1.75}$
in typical novae such as LV Vul (Figure \ref{optical_m10_abs_lv_vul_mag}b). 

The $V$ light curve of V339 Del does not smoothly decay along
the $\propto t^{-1.75}$ law, but has a $\sim 1.5$ mag dip in the
nebular phase.  We try to elucidate the nature of this $V$ dip
in Section \ref{discussion}.

\subsection{[\ion{O}{3}] band flux}
\label{o3_flux}

Figure \ref{optical_m10_abs_lv_vul_mag}a
shows the temporal variation of the [\ion{O}{3}] band flux, the data of
which are taken from \citet{mun15mm}.  Note that the day zero in their
Figures 5, 6, 7, and 8 are defined at the first optical peak, HJD 2,456,520.9,
which corresponds to our day 2.9.  
It decays almost along with the $V$ and $y$ light curves until the dust
blackout starts (gray-shaded region).
This simply means that the continuum flux dominates the line emission
in the [\ion{O}{3}] band.  If we divide the [\ion{O}{3}] band flux
into two parts, i.e.,
\begin{equation}
F_{\nu,\rm [O\,III]} = F_{\nu, \rm [O\,III], cont}
+ F_{\nu, \rm [O\,III], line},
\end{equation}
where $\nu$ is the frequency, $F_{\nu, \rm [O\,III], cont}$ is
the continuum flux in the [O\,III] band, and $F_{\nu, \rm [O\,III], line}$
is the [\ion{O}{3}] line flux.
This can be seen in the spectral evolution in Figure 4 of \citet{mun15mm}
that shows the emission line [\ion{O}{3}] is weak until day 9.4 and
gradually become strong and dominate the continuum on day 59.
Thus, we have $F_{\nu, \rm [O\,III], cont} \gg F_{\nu, \rm [O\,III], line}$ 
in the early phase, i.e., 
$F_{\nu, \rm ff, wind} \approx F_{\nu, \rm [O\,III], cont}$,
where $F_{\nu, \rm ff, wind}$ is the free-free flux of wind
in Equation (\ref{free-free_flux_v-band}).
This is the reason that the [\ion{O}{3}] band light curve follows
almost the free-free light curve.

After the dust blackout starts, the total continuum
flux of wind (FF+BB) decays quickly as indicated by the thin black line
in Figure \ref{optical_m10_abs_lv_vul_mag}a,
but the [\ion{O}{3}] band flux follows 
%the universal decline law of $L_{\rm [O\,III]} \propto t^{-1.75}$
%(thick cyan line labeled $t^{-1.75}$ or thick red line labeled shell).   
%Its decay behavior looks like that of 
the shell emission (thick red line labeled shell).
Now, we have $F_{\nu, \rm shell} \approx F_{\nu, \rm [O\,III], line}$,
%because the continuum flux (FF+BB) decays quickly,
where $F_{\nu, \rm shell}$ is the flux of the shocked shell
in Equation (\ref{luminosity_shocked_shell_ff_flux}).
This indicates that, after the dust blackout starts (in the
gray-shaded region), the [\ion{O}{3}] band flux is mainly originated
from the shocked shell but not from the wind.
%This is the reason why the [\ion{O}{3}] band flux does not experience
%a dust blackout, because the emission region is outside the dust shell.

After the dust blackout ended (outside the gray-shaded region),
the radiation field in the shocked shell becomes hot, so
the line flux, $L_{\rm [O\,III], line}$, increases.
As a result, the [\ion{O}{3}] band flux starts to deviate from the universal
decline law (thick cyan line labeled $t^{-1.75}$
%%%, i.e., we have $L_{\rm [O\,III], cont} \ll L_{\rm [O\,III], line}$
in Figure \ref{optical_m10_abs_lv_vul_mag}a).
Figure 4 of \citet{mun15mm} shows that the line flux dominates 
the [\ion{O}{3}] band flux on day 492. 
We regard that the [\ion{O}{3}] line emission is located
at the shocked shell, as shown in Figure \ref{wind_shock_config}d.
%(we may stress to say here, instead of writing as an item in conclusion, 
%so that Munari can easy to cite this paper.) 

\subsection{H$\alpha$ band flux}
\label{h_alpha_flux}

\begin{figure*}
%\epsscale{1.15}
\epsscale{0.5}
\epsscale{0.75}
\plotone{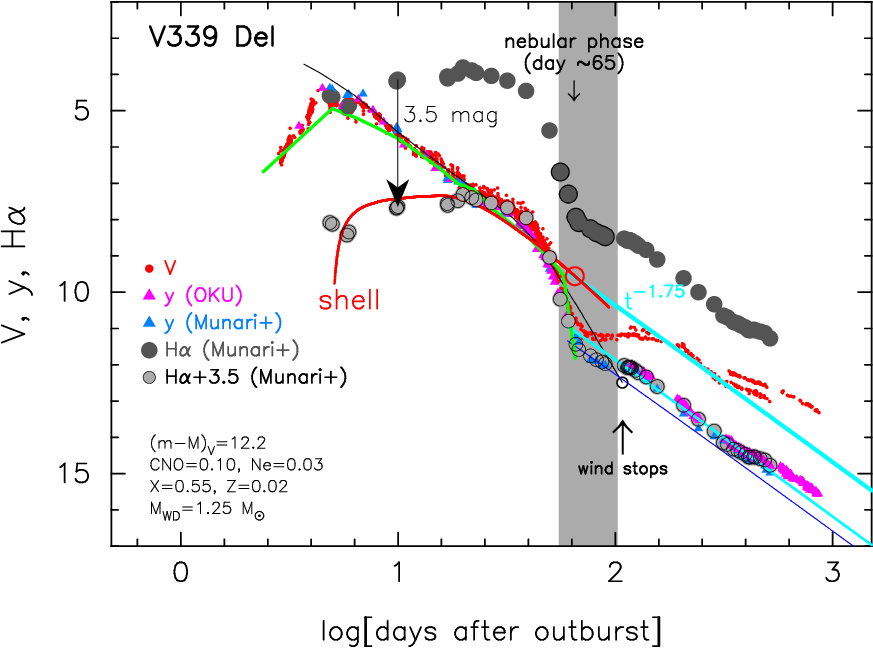}
%\rotate
%\plotone{f4.eps}
%\plotone{v339_del_m10_shell_big_y_o3_logscale_no3.epsi}
%\plotone{v339_del_lv_vul_m10_shell_logscale_no3.epsi}
%\plotfiddle{evolution1.ps}{5.0cm}{270}{0.4}{0.4}{-170}{220}
\caption{
Same as Figure \ref{optical_m10_abs_lv_vul_mag}a, but we add the
H$\alpha$ band light curve (dark-gray circles) instead of the [\ion{O}{3}]
band light curve.  The H$\alpha$ data are taken from \citet{mun15mm}.  
The encircled light-gray circles indicate their 3.5-mag-down data.  
\label{v339_del_m10_shell_big_y_o3_logscale_no3}}
\end{figure*}

Figure \ref{v339_del_m10_shell_big_y_o3_logscale_no3} shows the 
temporal variation of the H$\alpha$ band flux (filled dark gray circles),
the data of which are taken from \citet{mun15mm}. 
At/near the optical maximum, H$\alpha$ band magnitudes almost coincide with
those of $V$ and $y$ magnitudes, and after that it becomes much brighter than 
the optical $V$ and $y$ magnitudes. 

We add a 3.5 mag-down H$\alpha$ light curve (encircled light grey circles),
which shows an excellent agreement  
with the shell emission light curve (red line labeled shell)
until the dust blackout started and then with the $y$ light curve.

If we divide the H$\alpha$ band flux into two parts, i.e.,
\begin{equation}
F_{\nu, {\rm H}\alpha} = F_{\nu, {\rm H}\alpha,{\rm cont}}
+ F_{\nu, {\rm H}\alpha, {\rm line}},
\end{equation}
where $F_{\nu, {\rm H}\alpha,{\rm cont}}$ is the continuum flux 
in the H$\alpha$ band, and $F_{\nu, {\rm H}\alpha, {\rm line}}$
is the H$\alpha$ line flux at the frequency $\nu$,
we have $F_{\nu, \rm ff, wind} \approx 
F_{\nu, {\rm H}\alpha, {\rm cont}} \gg L_{\nu, {\rm H}\alpha, {\rm line}}$
at/near the optical maximum.  This is the same situation as that of
the [\ion{O}{3}] band flux in the previous subsection \ref{o3_flux}.

It is clear that the H$\alpha$ band flux follows proportionally
with the flux of shell emission, i.e., we have $F_{\nu, \rm shell}(t) 
\propto F_{\nu, {\rm H}\alpha, {\rm line}}(t) \gg F_{\nu, {\rm H}\alpha,
{\rm cont}}(t)$ after the strong shock arises and until the dust
blackout starts.  Thus, we conclude that the dominant source of 
H$\alpha$ is the shocked shell, at least, between the global optical
maximum and the start of dust blackout.
%In this case, the shell emission is dominated by bound-bound (H$\alpha$)
%emission and the coefficient in Equation 
%(\ref{luminosity_shocked_shell_ff_flux}) is $B_{\rm bb}$ for H$\alpha$.

The H$\alpha$ band flux rapidly decreases in the dust blackout phase.
We regard that the temperature of radiation field in the shocked shell
rapidly decreases with the dust formation. 
However, it is very interesting that the 3.5-mag-down H$\alpha$
light curve is just overlapped with the $y$ light curve after the dust
blackout has started.

%Fig.9

\subsection{Distance Modulus based on the Time-stretching Method}
\label{v339_del_time_stretch}

Here, we derive the distance modulus
to V339 Del based on the time-stretching method   
\citep{hac10k, hac15k, hac16k, hac18k,hac20skhs}. 
This method is based on the similarity of nova light curves.
Adopting appropriate time-stretching parameters, we are able to overlap
two nova light curves even if the two nova speed classes are different. 
If the two nova $V$ light curves, i.e.,
one is called the template and the other is called the target,
$(m[t])_{V,\rm target}$ and $(m[t])_{V,\rm template}$
overlap each other after time-stretching of a factor of $f_{\rm s}$
in the horizontal direction and shifting vertically down by $\Delta V$, i.e.,
\begin{equation}
(m[t])_{V,\rm target} = \left((m[t \times f_{\rm s}])_V
+ \Delta V\right)_{\rm template},
\label{overlap_brigheness}
\end{equation}
their distance moduli in the $V$ band satisfy
\begin{eqnarray}
(m-M)_{V,\rm target} 
&=& \left( (m-M)_V + \Delta V\right)_{\rm template} \cr
& & {~~~~} -2.5 \log f_{\rm s}.
\label{distance_modulus_formula}
\end{eqnarray}
Here, $m_V$ and $M_V$ are the apparent and absolute $V$ magnitudes,
and $(m-M)_{V, \rm target}$ and $(m-M)_{V, \rm template}$ are
the distance moduli in the $V$ band
of the target and template novae, respectively.

Figure \ref{optical_m10_abs_lv_vul_mag}b shows
the $V$ light curves of LV Vul and V339 Del.
%%% , c, and d 
%(c) $V$ light curves of V1668 Cyg and V339 Del, and
%(d) $V$ light curves of V1500 Cyg and V339 Del.
With the time-stretching factor and $V$ shift in 
Figure \ref{optical_m10_abs_lv_vul_mag}b,
we have the relation of
\begin{eqnarray}
(m&-&M)_{V, \rm V339~Del} \cr
&=& (m - M + \Delta V)_{V, \rm LV~Vul} - 2.5 \log 0.48 \cr
&=& 11.85 - 0.5\pm0.2 + 0.8 = 12.15\pm0.2, 
%\cr
%&=& (m - M + \Delta V)_{V, \rm V1668~Cyg} - 2.5 \log 0.48 \cr 
%&=& 14.6 - 3.2\pm0.2 + 0.8 = 12.2\pm0.2 \cr
%&=& (m - M + \Delta V)_{V, \rm V1500~Cyg} - 2.5 \log 0.91 \cr
%&=& 12.2 - 0.1\pm0.2 + 0.1 = 12.2\pm0.2,
\label{distance_modulus_v339_del_lv_vul_v1974_cyg_v_bv_ub}
\end{eqnarray}
where we adopt $(m-M)_{V, \rm LV~Vul}=11.85$ 
%$(m-M)_{V, \rm V1668~Cyg}=14.6$, and $(m-M)_{V, \rm V1500~Cyg}=12.2$
from \citet{hac21k}.
Thus, we have a consistent value of the distance modulus in the $V$ band
with the Gaia eDR3 distance of $d=2.1$ kpc and extinction $E(B-V)= 0.18$
mentioned in Section \ref{extinction_distance}.
The distance of LV Vul
%, V1668 Cyg, and V1500 Cyg are
is calculated to be $d=1.04\pm 0.1$ kpc for $E(B-V)=0.57$ \citep{schaefer22b}
and $(m-M)_{V, \rm LV~Vul}=11.85$ \citep{hac21k}.
%, $d=5.4\pm 0.5$ kpc for $E(B-V)=0.30$,
%and $d=1.4\pm 0.1$ kpc for $E(B-V)=0.45$, respectively.
This distance is consistent with the results of Gaia eDR3
\citep{bai21rf}, $d=1.17^{+0.2}_{-0.13}$ kpc.
% (LV Vul), 
%$d=3.7^{+2.1}_{-1.4}$ kpc (V1668 Cyg), 
%and $d=1.57^{+0.27}_{-0.19}$ kpc (V1500 Cyg).
Thus, our distance modulus in the $V$ band of $(m-M)_V=12.2\pm 0.2$ 
is supported by the time-stretching method.

%fig.10
%\placefigure{v339_del_m10_shell_big_logscale_no3}

\begin{figure*}
%%\epsscale{0.75}
%%\epsscale{0.8}
%%\epsscale{1.0}
\epsscale{1.15}
\plotone{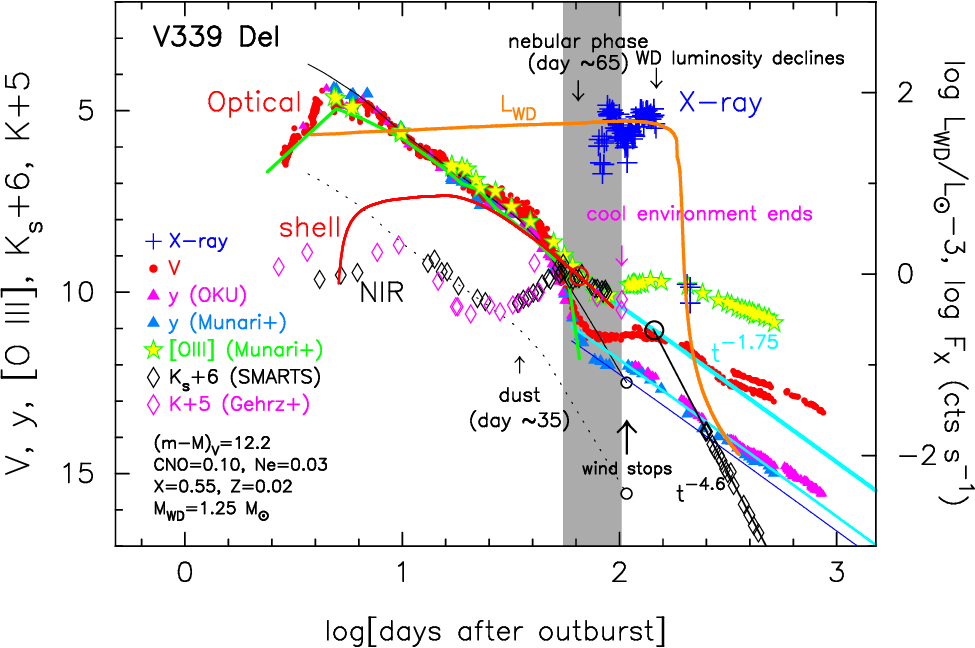}
%\plotone{v339_del_m10_shell_big_logscale_no3.epsi}
%\plotfiddle{evolution1.ps}{5.0cm}{270}{0.4}{0.4}{-170}{220}
\caption{
Same as those in Figure \ref{optical_m10_abs_lv_vul_mag}a,
but we add the $K_{\rm s}+6$, $K+5$ mag, and X-ray light curves of V339 Del.
The $K_{\rm s}$ data are taken from SMARTS \citep{wal12bt}.
The $K$ data are taken from \citet{geh15eh}.
Here, we plot both the $y$ magnitudes observed at OKU (filled magenta
triangles) and those taken from \citet{mun15mm} (filled cyan-blue triangles).
The thin dotted black line represents the same line of the 1.25 $M_\sun$
WD (Ne2) model light curve (thin solid black line),
but is shifted down by 3 mag to fit it with the observed 
$K_{\rm s}+6$ light curve in the early decay phase (see
Figure \ref{v339_del_v_bv_ub_color_curve_no2}).  
The thick straight black line labeled $t^{-4.6}$ denotes the very late
decline trend of $L_{K_{\rm s}} \propto t^{-4.6}$, a simple extension of
which crosses the universal decline trend (thick cyan line)
at day $\sim$140 (large open black circle).
The thin straight cyan line has the same decline rate as that of
the universal decline law, but 1.5 mag below.  This line represents the trend
of the $y$ magnitude.  The thin blue line is shifted further down by 0.4 mag.
The gray-shaded region represents the substantial
absorption phase by a dust shell.
We also added the photospheric luminosity $L_{\rm WD}= L_{\rm ph}$ of
the WD by the thick orange line.  The arrow labeled ``WD luminosity declines''
means the date when the WD luminosity starts to decline, on day $\sim 140$. 
\label{v339_del_m10_shell_big_logscale_no3}}
\end{figure*}

\section{Discussion}
\label{discussion}

V339 Del is a rare object among novae in the sense that dust exists 
in the SSS phase \citep{geh15eh}.
This means that a hot radiation field did not destroy dust. 
In this section, we clarify the reason why a dust shell 
coexists with a soft X-ray radiation field in V339 Del,
and confirm that the 1.5 mag dip in the $V$ light curve is    
not caused by dust absorption.

\subsection{[\ion{O}{3}] emission and rapid decay of the $V$ light curve}
\label{rapid_decay_vy}

In Section \ref{v_light_curve_fitting}, we have shown that 
the rapid drop of the $y$ magnitude is due to the rapid decrease
in the wind mass-loss rate, not by the dust absorption.  
Such a drop in the $y$ magnitude was reported in
the two classical novae, V1500 Cyg \citep{loc76m} and V1668 Cyg
\citep{gal80ko}.  In marked contrast to V339 Del,
the $V$ magnitudes in these two novae display continuous decreasing
along with the universal decline law of $L_V\propto t^{-1.75}$.
This is because the flux of [\ion{O}{3}] rapidly increases and fills
the difference between the $y$ magnitude and $L_V\propto t^{-1.75}$ line 
\citep{hac16k}, as discussed in Section \ref{o3_flux}.

[\ion{O}{3}] lines are excited in an environment of lower density 
and higher temperature (hot radiation field).
In the rapid drop phase of free-free (FF)
light curve, the wind mass-loss rate quickly decreases that results
in the rapid drop of the wind density.  At the same time, the photospheric
temperature increases with time (Figure \ref{v339_del_v_logscale_no3}).
Therefore, the ejecta satisfy a suitable condition for exciting
[\ion{O}{3}] emission, as the mass-loss rate rapidly decreases. 

The increase in the [\ion{O}{3}] flux compensates the decrease in
the $y$ flux, which results in a continuous decline trend in the $V$
band luminosity expressed by $L_V\propto t^{-1.75}$ 
in typical novae such as V1500 Cyg and
V1668 Cyg.  In V339 Del, however, dust blackout phase overlaps with
this rapid drop phase of $V,y$ light curves, as shown in Figure 
\ref{v339_del_m10_shell_big_logscale_no3}.  The [\ion{O}{3}] lines are
not effectively excited in a cooler environment of dust formation
(Figure \ref{wind_shock_config}d).
Therefore, the gap is not filled with the [\ion{O}{3}] flux
until a hot radiation environment is recovered.
We can interpret that the 1.5 mag dip is caused by a cooler environment 
for dust formation but not by absorption of dust.
This interpretation is supported by the fact that the $y$ light curve
is not recovered just after the 1.5 mag dip but is monotonically
decreasing with the slope of $L_y\propto t^{-1.75}$ even after the dust
blackout phase ended (Figure \ref{v339_del_m10_shell_big_logscale_no3}).
Thus, the cool environment for dust formation affects the excitation
of [\ion{O}{3}] lines in the shocked shell (Figure \ref{wind_shock_config}d).

The $V$ band light curve shows a deeper dip of $\sim 1.5$ mag, 
as seen in Figure \ref{v339_del_m10_shell_big_logscale_no3}.   
This 1.5 mag depth corresponds exactly to the difference between 
the two cyan lines of $\propto t^{-1.75}$ for the $V$ and $y$ magnitudes. 
The $V$ band flux includes the flux of the [\ion{O}{3}] lines while
the $y$ band does not.  After the dust blackout phase (gray-shaded region
in Figure \ref{v339_del_m10_shell_big_logscale_no3}), the [\ion{O}{3}] 
emission rapidly increases.  This increase begins when a cool environment
disappears and a hot radiation {\bf field} is recovered.
Thus, the deep depth of 
1.5 mag in the $V$ band simply suggests that the optical line emission
of [\ion{O}{3}] is substantially suppressed during the dust blackout phase,
and not owing to absorption of continuum flux 
by dust.

\subsection{Formation of an optically-thin dust shell}
\label{thin_dust_shell}

Figure \ref{v339_del_m10_shell_big_logscale_no3} shows 
the near-infrared (NIR) $K_{\rm s}$ and $K$ band magnitudes, but 
$K_{\rm s}$ mag is shift down by 6 mag and $K$ mag is 5 mag down
to fit them with the universal decline trend of $\propto t^{-1.75}$
(thick cyan line labeled $t^{-1.75}$) during the dust blackout phase. 
The $K_{\rm s}+6$ shows two maxima, on day $\sim 11$ and day $\sim 55$. 
After the first maximum, the $K_{\rm s}+6$ magnitude declines 
along with our $1.25 ~M_\sun$ WD (Ne2) model light curve
(dotted black line), same as the thin black line, but 
shifted down by 3 mag, to fit the $K_{\rm s}+6$ mag data. 
Because this line reproduces well the observed NIR data before rebrightening,
we regard that the NIR emission in this phase is dominated by free-free 
emission from the optically thin plasma (winds) just outside the photosphere. 

The $K_{\rm s}$ magnitude starts to rise again after a dust shell forms
on day $\sim 35$, and reaches the second peak around day $\sim 55$.
After that, the $K_{\rm s}$ magnitude begins to decay following
the universal decline law of $L_{K_{\rm s}}\propto t^{-1.75}$, where
$L_{K_{\rm s}}$ is the luminosity in the $K_{\rm s}$ band.
The $K+5$ magnitude also behaves similarly to the $K_{\rm s}+6$ magnitude
until a substantial dust blackout phase ended (shaded region in Figure
\ref{v339_del_m10_shell_big_logscale_no3}).

%%xxxxxxxxxxxxxxxxxxxxxxxxxx

\citet{der17ml} showed that dust formation can occur efficiently within
the post-shock gas.  Therefore, we regard that a dust shell formed
just behind the shock as illustrated in Figure \ref{wind_shock_config}d.
Once the dust shell formed, it partly blocks FF+BB (thin black
line in Figure \ref{v339_del_m10_shell_big_logscale_no3}) emissions 
from the wind and WD photosphere, which are emerging from near WD, i.e.,
far inside the shock and dust shell.  
Taking a close look at Figure \ref{v339_del_m10_shell_big_logscale_no3},
we can see that the $y$ band radiation is slightly absorbed between 
day 75 -- 90.  During this period, it decays along with the thin blue line of  
$\propto t^{-1.75}$, which is 0.4 mag below the thin cyan line.
If the $y$ band fluxes were not absorbed by the dust shell, 
the $y$ light curve would follow this thin cyan line.
Thus, we measure the optical depth of dust at the $y$ band;
the $y$ band flux was absorbed, at least,
by 0.3 mag $= 10^{-0.3/2.5} = 0.76 = e^{-0.28}$, that is, $\tau_y= 0.28$, 
or 0.4 mag $= 10^{-0.4/2.5} = 0.69 = e^{-0.37}$, i.e., $\tau_y= 0.37$.  

A similar analysis can be done in the UV light curves.  \citet{sho18km}
showed three UV light curves of 2025, 2650, and 3555\AA\ 
bands in their Figure 11.  They fitted the overall decay trend with
$L_{\rm UV}\propto t^{-1.5}$, where $L_{\rm UV}$ is the luminosity
of a UV band.  These three UV light curves show a dip between day 80
and 100.  Each measured flux data shows large fluctuations.
The 3555\AA\  band data are least scattered among the three bands,
so we adopted the 3555\AA\  light curve and estimated the deepest depth
of the dust dip, as deep as $2.25/3.0 \approx e^{-0.3}$ on day 73,
the difference from the fitted trend line of $L_{\rm UV}\propto t^{-1.5}$.  
This corresponds to the optical depth of $\tau_{\rm 3555\AA }= 0.3$, being
consistent with the dust absorption calculated from the $y$ band light
curve as mentioned above.

% 2,456,520.5  = 2013 August 16.0
% August 15.94 = 520.44  --> B5
% August 16.86 = 521.36  --> A0  at/near optical maximum

% dust formation
% 557.28 (39.28 days) 6.5E13 cm    4E3 Lo  8E-10 Mo  
% 577.18 (59.18 days) 1.7E14 cm  1.1E4 Lo  5E-9  Mo
% T_ph=5.07 --> L= dex(38.25) = 1.8E38 erg/s = 4.6E4 Lo
%  

\citet{tar14tst} estimated the dust luminosity of
$L_{\rm dust}\sim 4\times 10^3 (d/{3 \rm ~kpc})^2 ~L_\sun$ and
$\sim 1.1\times 10^4 (d/{3 \rm ~kpc})^2 ~L_\sun$ on day 39.28
and 59.18, respectively.  The latter luminosity corresponds to 
10\% of the WD luminosity ($4.6\times 10^4 ~L_\sun$ for our 1.25 $M_\sun$
WD model together with the distance of $d=2.1$ kpc).  In other words,
the dust shell absorbed $\sim 10$\% of UV and soft X-ray photons
from the central hot WD, and reemitted in near/mid-infrared bands.
This picture is consistent with our interpretation of optically-thin dust
shell, $\tau \sim 0.3$ estimated in the $y$ band.

%%% JD 2456555.5 - 518.0 = 37.5+0.5 days
\citet{sko14dt} presented a dust component in their broad band spectrum 
on day 38 (UT 2013 September 20).  The dust 
% component has a blackbody
temperature is $T_{\rm dust}= 1350\pm 50$ K and the luminosity is 
$L_{\rm dust}= (1.1\pm 0.2)\times 10^{37}$ erg s$^{-1}(d/{\rm 3~kpc})^2$ 
with blackbody assumption.  
This dust luminosity is about 3\% of the central WD luminosity,
i.e., $L_{\rm dust}/L_{\rm WD}= 5.4\times 10^{36}/ 1.6\times 10^{38} = 0.03$
on day 38 of our $1.25 ~M_\sun$ WD model and distance of $d=2.1$ kpc.
Assuming that dust absorbs a part of the flux from the central WD and
reemits it and the total emitting flux is conserved (steady-state),
the absorption efficiency by dust is about 3\%.  This corresponds to the
optical depth of $\tau= 0.03$.  The shock luminosity is negligibly small
on day 38, i.e., $L_{\rm sh}= 1.6\times 10^{35}$ erg s$^{-1}$
from Equation (\ref{shocked_energy_generation})
with $\dot{M}_{\rm wind}= 5.7\times 10^{-6} ~M_\sun$ yr$^{-1}$.
%The formation date of dust is broadly consistent with the date reported by
%\citet{shen13tt}.

\citet{tar14tst} also estimated the radius of the dust shell, 
$R_{\rm dust}\sim 6.5\times 10^{13} ~(d/{3 \rm ~kpc})$ cm and  
$R_{\rm dust}\sim 1.7\times 10^{14} ~(d/{3 \rm ~kpc})$ cm on day 39.28
and 59.18, respectively.  
%\\{\bf (instead of the following part I wrote the red sentences) \\
%The position of our shocked shell is
%about $R_{\rm shock} = 2.7\times 10^{14}$ cm on day 59, which is twice
%larger than \citet{tar14tst}'s estimate.  Assuming an optically-thick
%sphere of dust shell as a radiation source, they obtained 
%the radius of the dust shell from $R_{\rm dust}= D\times
%(F(L)/B(L,T_{\rm C}))^{0.5}$, where $D$ is the distance, $F(L)$ is the
%luminosity at 3.5 $\mu$m ($L$ band), and $B(L,T_{\rm C})$ is the 
%intensity of the blackbody radiation at the $L$ band and dust temperature
%of $T_{\rm C}$.  However, we must adopt the filling factor of 
%$f= h/R_{\rm dust}$ when the dust shell is optically thin.
%}
On the other hand, our shocked shell is located 
at $R_{\rm shock} = 2.7\times 10^{14}$ cm on day 59, 
twice larger than \citet{tar14tst}'s estimate for $d=2.1$ kpc.  
This difference can be explained as follows:
Assuming an optically-thick
sphere of dust shell as a radiation source,  
\citet{tar14tst} estimated the dust shell radius, 
as $R_{\rm dust}= D\times (F(L)/B(L,T_{\rm C}))^{0.5}$, 
where $D$ is the distance, $F(L)$ is the luminosity 
at 3.5 $\mu$m ($L$ band), and $B(L,T_{\rm C})$ is the 
intensity of the blackbody radiation at the $L$ band and dust temperature
of $T_{\rm C}$. 
%This estimate is based on the assumption that 
%a radiation source is an uniformly emitting disk in the sky of 
%the filling factor of one. 
In our analysis, the dust shell is optically thin (and geometrically-thin)
so that we must adopt a filling factor, $f=h/R_{\rm dust}$, where $h$ is
the thickness of the dust shell.  Then, the total emission 
of the dust should be written as $4\pi R_{\rm dust}^2 (h/R_{\rm dust}) \times
B(L,T_{\rm C}) = 4\pi D^2 \times F(L)$.
%, where $h$ is the thickness of the dust shell.
If we adopt $h=0.1 ~R_{\rm dust}$ ($f=0.1$) or so, we obtain
$R_{\rm dust}^2= 10\times D^2\times
F(L)/B(L,T_{\rm C})$, which gives a radius of three times larger than
their estimate, being broadly consistent with the shock
radius of our model for the distance of $d=2.1$ kpc.    

\citet{geh15eh} reported that the dust shell emitted about 
$L_{\rm dust}\sim 0.02 ~L_o = 3.6\times 10^3 ~L_\sun$ on day 102,
which is about 8\% of the luminosity of our 1.25 $M_\sun$ WD model.
Here, $L_o= 1.8\times 10^5 ~L_\sun$ for the distance of $d=2.1$ kpc.
This dust luminosity is consistent with \citet{tar14tst}'s estimate.

\citet{geh15eh} also argued that the failure of substantial dust formation
is due to a lower gas density than the critical density proposed by
\citet{geh87n}.

To summarize, during the substantial dust blackout phase 
(shaded region in Figure \ref{v339_del_m10_shell_big_logscale_no3}),
a dust shell with the optical depth of $\tau= 0.3-0.4$ absorbed  
the $y$ band flux (FF+BB: continuum) that comes from far inside the 
shocked shell. 
 
%xxxxxxxxxxxxxxxxxxxxxxxxxxxxxxx

%Fig.11  
%\placefigure{wind_shock_config}

\begin{figure*}
\epsscale{1.0}
\plotone{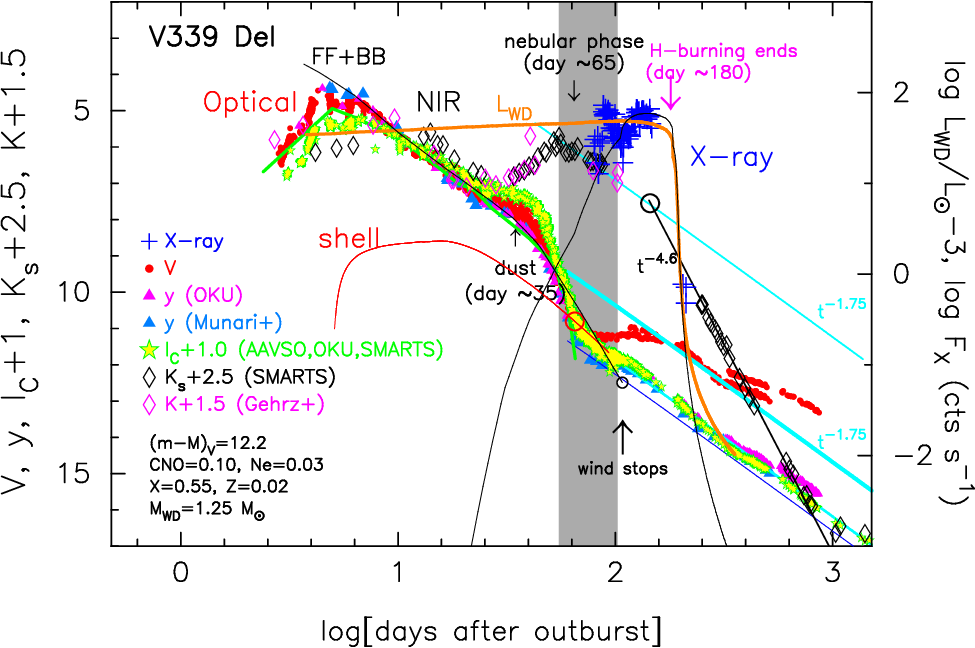}
%\plottwo{f1.eps}{f2.eps}
%%%%\plotone{mdot_radius_velocity.epsi}
%%%%\plotone{opt_gamma_ray_fluxes.epsi}
%\plotfiddle{evolution1.ps}{5.0cm}{270}{0.4}{0.4}{-170}{220}
\caption{
Same as Figure \ref{v339_del_m10_shell_big_logscale_no3}, but the
$I_{\rm C}+1$ mag light curve is added.  The $I_{\rm C}$ data are taken
from AAVSO, OKU, and SMARTS.  We also add the bolometric luminosity of
$L_{\rm WD}$ ($\equiv L_{\rm ph}$: orange line labeled $L_{\rm WD}$)
and X-ray (0.3--2.0 keV) model light curve (thin black line)
of our $1.25 ~M_\sun$ WD (Ne2),
which is the same as that in Figure \ref{v339_del_v_logscale_no3}.
The $I_{\rm C}+1$ magnitude light curve broadly follows a slope of
$L_{I_{\rm C}}\propto t^{-1.75}$ in the early and late phase but
the later phase light curve is about 1.5 mag below that of the early one.
We also plot the temporal variation of shell emission (shocked shell) 
by the thin red line, which is the same as the thick red line in
Figure \ref{v339_del_m10_shell_big_logscale_no3}, but 1.5 mag down. 
\label{v339del_vs_v5668sgr}}
\end{figure*}

\subsection{The 1.5 mag dip in the $V$ band}
\label{dip_v_mag}

We showed that the dust is optically thin in Section \ref{thin_dust_shell}.
If the 1.5 mag dip in the $V$ light curve is attributed to dust absorption,
we need much larger optical depth such as $\tau \sim 1.4$, which is unlikely 
to be compatible with the observation as described in Section 
\ref{thin_dust_shell}.
Thus, we conclude that the 1.5 mag dip is not caused by absorption of dust.

From Figure \ref{v339_del_m10_shell_big_logscale_no3},
%%%\ref{v339del_vs_v5668sgr} 
we see that (1) the temporal variation of the [\ion{O}{3}] flux behaves
similarly to the shell emission during the substantial dust blackout phase 
(gray-shaded region in Figure \ref{v339_del_m10_shell_big_logscale_no3}).
Thus, the [\ion{O}{3}] emission region must be in the shocked layer.
(2) It is unlikely that the flux of [\ion{O}{3}] lines is much absorbed
during the dust blackout phase.
Instead, it almost follows the universal decline law.
Thus, we suppose that these [\ion{O}{3}] lines are formed outside the 
optically-thin dust layer as shown in Figure \ref{wind_shock_config}d.
(3) The dust formation site needs a cool environment. 
The resultant cool environment could suppress the excitation of
[\ion{O}{3}] lines.
(4) Dust formation reaches a peak on day 55.  After day 55, the cool
environment is gradually diluted out.  A hot radiation could start to excite
[\ion{O}{3}] lines in the dust layer at least until day 100,
the end of substantial dust blackout phase.
(5) The flux of [\ion{O}{3}] lines gradually increases and reaches a peak
on day 140.  Accordingly, the $V$ light curve recovers and follows
the original universal decline law (thick cyan line labeled $t^{-1.75}$
in Figure \ref{v339_del_m10_shell_big_logscale_no3}).

The 1.5 mag drop in the $V$ band simply suggests that excitations
of [\ion{O}{3}] lines are substantially suppressed during the dust blackout
phase, but it is not owing to large absorption by dust.

\subsection{Rapid decrease in the dust shell emission}
\label{disappearence_dust_shell}

Although we have no data between day $\sim 103$ and day $\sim 248$, 
we see a clear change in the decline rate of $K_{\rm s}+2.5$ (or
$K+1.5$) light curve (open black (or magenta) diamonds) in 
Figure \ref{v339del_vs_v5668sgr};
it decays along the universal decline law of $\propto ~t^{-1.75}$
(thin cyan line labeled $t^{-1.75}$) in the dust blackout phase,
but in the latest phase decays as $L_{K_{\rm s}} \propto t^{-4.6}$
(thick straight black line labeled $t^{-4.6}$).
A simple extension of the black straight line can cross the universal
decline law (thin cyan line) on day $\sim 140$ (large open black circle).
This date is close to the date when the WD luminosity starts to decline
on day $\sim 140$ (thick orange line labeled $L_{\rm WD}$ in 
Figures \ref{v339_del_m10_shell_big_logscale_no3} and 
\ref{v339del_vs_v5668sgr}).

The decay trend of $L_{K_{\rm s}}\propto t^{-1.75}$ could be
explained by the relative decrease in the cross section of dust.
We assume that the number of dust grains and their size are unchanged
after the maximum of $L_{K_{\rm s}}$ on day 55 
(Figure \ref{v339del_vs_v5668sgr}).
As the dust shell expands with the shocked shell,
the surface area of the shell increases as 
$A_{\rm dust}= 4\pi R_{\rm dust}^2= 4\pi (v_{\rm shock} t)^2$, where
$R_{\rm dust}$ is the radius of the shell, $t$ is the time from
the formation of a shock, and $R_{\rm dust}\approx v_{\rm shock} t$.  
On the other hand, the total cross section of dust grains is
constant with time, i.e., $S_{\rm dust}=$constant because the dust shell 
is optically thin.
The photon cross section against dust is decreasing
along $S_{\rm dust}/A_{\rm dust}\propto t^{-2}$, here the shock 
velocity is almost constant (very gradually increasing). 
The absorption by the dust shell decreases as $t^{-2}$ 
and the re-emission from the dust also decreases with the same rate $t^{-2}$. 
This decay trend is very close to the 
trend of $L_{K_{\rm s}}\propto t^{-1.75}$.

\citet{sho18km} discuss that, since the structures are frozen within
the ejecta, any change in IR continuum emission should be a simple power law.  
Thus, we may extend the decay trend of $L_{K_{\rm s}}\propto t^{-1.75}$
to day $\sim 140$, the date when the WD luminosity starts to decline
as shown in Figure \ref{v339_del_m10_shell_big_logscale_no3}, or
day $\sim 180$, the date when the hydrogen shell-burning ends
as shown in Figure \ref{v339del_vs_v5668sgr}.

After the Sun constraint, the decay trend was changed to 
$L_{K_{\rm s}} \propto t^{-4.6}$.  It is reasonable to 
assume that this trend went back to the crossing point with
the trend of $L_{K_{\rm s}}\propto t^{-1.75}$ (or
$L_{K_{\rm s}}\propto t^{-2}$).

This crossing point coincides with the date when the $V$ light curve
fully recovered from the 1.5 mag dust dip or when the [\ion{O}{3}]
light curve attains its local maximum (Figure 
\ref{v339_del_m10_shell_big_logscale_no3}).
Both of them could be related
to the local maximum of the WD luminosity, $L_{\rm WD}= L_{\rm ph}$,
where we define the WD luminosity as the photospheric luminosity
of our $1.25 ~M_\sun$ WD envelope (orange line in Figures
\ref{v339_del_m10_shell_big_logscale_no3} and \ref{v339del_vs_v5668sgr}).  
Except for the dust blackout phase, the $V$ light curve
broadly follows the $L_V \propto t^{-1.75}$ law
from its maximum until a very late phase. 

The decay trend of $L_{K_{\rm s}}$ could have changed later than the 
crossing point, if it is associated to 
a rapid cooling of the WD (day $\sim 180$ in our model). 
Or it could change earlier: 
\citet{eva17bg} reported that the dust flux, grain size,
and dust mass have peaked on day 103, 
although there are no data between day 103 and day 683.

\citet{sho18km} claimed that dust survives at least day 867 
because optical and UV line profiles show asymmetry 
owing to dust absorption. 
Asymmetry of IR line was observed until day 78 
but disappeared on day 683 \citep{eva17bg}, which 
Shore et al. explained as the decreased  
dust opacity, too low to cause asymmetry in IR.

\citet{sho18km} estimated the optical depth of dust, $\tau_{\rm d}\approx 1$
even after day 650. This value is too large 
because the $V$ light curve almost recovered 
on day 140 
(thick solid cyan line in Figure \ref{v339del_vs_v5668sgr}.
If $\tau_{\rm d}$ is as large as $\tau_{\rm d}\approx 1$, 
the $V$ light curve should be fainter by $-2.5 \log e^{-1} = 1.1$ mag.
The complete recovery to the overall universal decline law may support
our results that the absorption of $\tau=0.3-0.4$ 
had disappeared at least until day 140.

\subsection{Overall decline trends of $I_{\rm C}$ band}
\label{comparison_v5668_sgr}

The $I_{\rm C}+1$ mag light curve in Figure \ref{v339del_vs_v5668sgr} 
follows well the $y$ light curve. 
This is a remarked contrast with the $V$ light curve that brightens 
after the dust blackout, because $V$ band 
is heavily contributed by strong emission lines 
such as [\ion{O}{3}] lines. 
This means that the $I_{\rm C}$ band is less contributed 
by strong emission lines. 
A close look at the $I_{\rm C}$ light curve shows 
a small bump on day 35 - 90, which is due to a blueward tail of
dust emission \citep[e.g.,][]{sko14dt}.
Also, {\bf the lack of} 0.4 mag dip on day 70 - 100 compared
with the $y$ light
curve could be due to contribution of a tail of dust emission.  
We clearly show that the overall $I_{\rm C}$ light curve of V339 Del 
is essentially emission line-free and represents continuum flux like
the $y$ light curve.

We plot the shell emission light curve (thin red line) in
Figure \ref{v339del_vs_v5668sgr}, which is the same as in Figure
\ref{v339_del_m10_shell_big_logscale_no3} but shifted down by 1.5 mag
to fit with the $I_{\rm C}$ light curve in the later phase.
We regard that this line
%the brighter one in Figure \ref{v339_del_m10_shell_big_logscale_no3}
%represents the case of bound-bound emission, i.e., $B_{\rm bb}$,
%for [\ion{O}{3}] emission lines without suppression, but
corresponds to the free-free emission for $I_{\rm C}$ band,
i.e., Equation (\ref{luminosity_shocked_shell_ff_flux}) with $B_{\rm ff}$,
where $B_{\rm ff}$ is the coefficient for free-free emission.
Both the $y$ and $I_{\rm C}$ light curves
show a rapid $1.5-2.0$ mag drop between day 50 and 70, and then, again,
follow the universal decline trend of $\propto t^{-1.75}$.  Before day 80,
the dominant continuum flux is the free-free flux from the winds
(FF+BB: thin solid black line), but is replaced with
the free-free flux from the shocked shell (shell: thin solid red line)
after day 80.  The shell emission also follows the trend of
$\propto t^{-1.75}$ after day 80.  
Both the $y$ and $I_{\rm C}$ light curves did not recover like the $V$ band.
This clearly shows that the continuum fluxes of V339 Del are not heavily
absorbed by dust.  The $1.5-2.0$ mag drops in the $y$ and $I_{\rm C}$ bands
are caused by the drop (decrease) in their free-free fluxes of winds.

To summarize, with no contribution of [\ion{O}{3}] lines, 
the $V$ light curve behaves similarly to the $y$ and $I_{\rm C}$ 
light curves because the $y$ and $I_{\rm C}$ bands represent 
emission line-free (continuum) fluxes.  The 1.5 mag drop itself
in the $V$ magnitude is caused by a sharp drop in the wind mass-loss rate
as calculated with Equation (\ref{free-free_flux_v-band}) and plotted
in Figure \ref{v339_del_v_logscale_no3}.

\subsection{Summary: coexistence of X-ray and dust shell}
\label{summary_coexistence_x_dust}

From detailed analysis of $V$, $y$, [\ion{O}{3}], $I_{\rm C}$,
and $K_{\rm s}$ (and $K$) light curves together with our theoretical
light curves, we may conclude that an optically-thin dust shell forms
behind the shock until about 55 days ($K_{\rm s}$ peak) after the outburst. 
The drop in the $V$ band during day 55 -- 140 is estimated to be as deep
as about 1.5 mag, which is caused by the substantial suppression of
strong emission lines such as [\ion{O}{3}] 4959, 5007 \AA\  
\citep[e.g.,][]{mun15mm}.  This drop correspond to $\sim 70$\% of
the shell $V$ emission ([\ion{O}{3}]), which is converted to,
and reemitted in, the near-IR $KLM$ (or $K_{\rm s}$) energy bands.

From the very shallow dust blackout in the $y$ magnitude (FF+BB and shell), 
we estimated the optical depth $\tau$ of the dust shell to be 
$\tau \lesssim 0.3$--0.4. This depth of $\tau \lesssim 0.3$
is also confirmed in the UV 3555\AA\ band \citep{sho18km}.
Because the dust shell is optically thin, 
continuum radiation in the $y$ band can mostly penetrate 
and also supersoft X-ray photons are little absorbed. 
This is the reason why we have large X-ray flux during the dust dip
in the $V$ light curve. 

After the SSS phase started on day 72, dust grains could be destructed
\citep[e.g., shattering process due to
electrostatic stress after the dust is exposed
to X-ray radiation: ][]{eva17bg}, but we have found no evidence
for substantial dust destruction in our light curve analysis.
Thus, the coexistence of dust and X-ray radiation simply means that we
see the X-rays through an optically-thin dust shell.


%% Included in this acknowledgments section are examples of the
%% AASTeX hypertext markup commands. Use \url without the optional [HREF]
%% argument when you want to print the url directly in the text. Otherwise,
%% use either \url or \anchor, with the HREF as the first argument and the
%% text to be printed in the second.

\section{Conclusions}
\label{conclusions}
Our main results are summarized as follows:
\begin{enumerate}
\item We present $BVyI_{\rm C}$ photometry of V339 Del at 
Osaka Kyoiku University from JD 2,456,522.0 to JD 2,457,382.0 for
about 860 days. 
\item We obtain the distance modulus in the $V$ band  
to be $\mu_V\equiv (m-M)_V=12.2\pm0.2$, applying the time-stretching
method to the $V$ light curve of V339 Del. 
The distance is $d=2.1\pm0.2$~kpc for the reddening of $E(B-V)=0.18$, which
is consistent with the Gaia eDR3 distance of $d=2.06^{+1.22}_{-0.75}$~kpc
\citep{bai21rf}. 
\item Our $1.25 ~M_\sun$ WD (Ne2) model well reproduces the $V,y$ light
curves with $(m-M)_V=12.2$, as well as the supersoft X-ray (0.3-2.0 keV). 
Here our model $V$ light curve consists of free-free emission from the ejecta
just outside the photosphere plus blackbody emission from the WD photosphere
(FF+BB).  The X-ray light curve is calculated with blackbody emission
of the photosphere.
\item The position of V339 Del in the maximum magnitude versus rate of
decline (MMRD) diagrams calculated by \citet{hac20skhs} gives a set of
the WD mass $M_{\rm WD}\approx 1.25 ~M_\sun$, mass-accretion rate
$\dot{M}_{\rm acc}\approx 3\times 10^{-9} ~M_\sun$ yr$^{-1}$, and
recurrence time $t_{\rm rec}\sim$2,000 yr.  These values are consistent
with our $1.25 ~M_\sun$ WD (Ne2) model.  Thus, we are able to easily
estimate the nova properties from these MMRD diagrams.
\item The expansion parallax method cannot be applied to a nova fireball stage,
simply because, once winds begin to blow, winds leave the pseudo-photosphere
behind, and the expansion rate of the pseudo-photosphere (280 km s$^{-1}$) 
is much smaller than the velocities of ejecta (613 or 505 km s$^{-1}$).
For example, the distance estimate based on the expansion parallax method
gave a distance of $d=4.5$ kpc \citep{schaefer14bg,geh15eh}, which is much
larger than our estimate of $d=2.1\pm 0.2$ kpc.
\item We show that, based on our $1.25 ~M_\sun$ WD (Ne2) model,
a strong shock arises just after the optical 
maximum and it moves outward far outside the nova photosphere 
with the shock velocity of 500--600 km s$^{-1}$.  
\item The shock energy is calculated to be $L_{\rm sh}\sim 6.8\times 10^{36}$
erg s$^{-1}$ from Equation (\ref{shocked_energy_generation}).
The ratio of $L_{\gamma}/L_{\rm sh}\sim 0.01$ satisfies the theoretical
request \citep[$L_{\gamma}/L_{\rm sh} \lesssim 0.03$, ][]{met15fv}.
Here the observed GeV gamma-ray energy is $L_{\gamma}\sim 6.5\times 10^{34}$
erg s$^{-1}$ \citep{ack14aa} for $d=2.1$ kpc.
\item The GeV gamma-ray flux observed with the Fermi/LAT 
decays twice slower than the optical flux.  This tendency
can be explained based on our optically-thick wind theory;  
the free-free emission flux depends on 
$\propto {(\dot M}_{\rm wind}/v_{\rm wind})^2$, while the  
shocked energy generation rate depends roughly on
$\propto {\dot M}_{\rm wind}/v_{\rm wind}$. 
\item After the strong shock arises just after the optical maximum,
a hot radiation field accelerates emission of H$\alpha$ in the shocked shell.
The temporal variation of the H$\alpha$ band flux decays 
following the shocked shell emission of our model.
Thus, the dominant source of the H$\alpha$ band is the shocked
shell, at least, from the optical maximum to the dust blackout.
\item A dust shell begins to form just behind the shock about 
35 days after the outburst.  The optical depth of the dust shell
can be estimated from the 0.4 mag dip in the $y$ light curve
during day 60-100.  The dust absorption is as small as 
$\tau_y \lesssim 0.3$--0.4 in the $y$ continuum flux.
Both the FF+BB emissions and supersoft X-rays are hardly absorbed.
This is the reason why the dust emission and supersoft X-rays coexist.
%\item The overall $V$ light curve decays along with a single decline law
%of $L_V\propto t^{-1.75}$ except for the dust blackout phase.
%Before day 70, both the $V$ and $y$ light curves decay along with
%our theoretical light curve that calculated from free-free emission of
%plasma (winds) outside the photosphere. 
%After day 140, the $V$ band is dominated by [\ion{O}{3}] line emission,
%and the $V$ light curve comes back to the single decline law of
%$L_V\propto t^{-1.75}$. 
\item The 1.5 mag drop in the $V$ light curve is caused not by dust
absorption but by a rapid decrease in the free-free flux from the nova wind.
This 1.5 mag drop in continuum flux is confirmed in the $y$ as well as
$I_{\rm C}$ bands. 
\item Dust forms in a cool and dense shell just behind the radiative shock
\citep{der17ml}.  Such a cool environment in the dust shell substantially
suppresses excitation of strong emission lines [\ion{O}{3}] 
in the very early nebular phase, which prevents recovery from a
1.5 mag $V$ drop on day 65.
\item The recovery from the 1.5 mag drop in the $V$ light curve
is caused by the delayed increase in the [\ion{O}{3}] line flux.
This is confirmed by the fact that
the $y$ band is essentially line-free (continuum) and the $y$ light curve
does not show any recoveries from the 1.5 mag drop as well as the
$I_{\rm C}$ light curve.
\item We have found no clear evidence for substantial dust destruction
in our light curve analysis.  
The coexistence of dust and X-ray radiation simply means that we
see X-rays through an optically-thin dust shell.  
\end{enumerate}

The $BVyR_{\rm C}I_{\rm C}$ data of V339 Del observed at OKU will be shared
on reasonable request to the authors.

\begin{acknowledgments}
We are grateful to the anonymous referee for useful comments
regarding how to improve the manuscript.
%     We are grateful to T. Kato for sending us the VSNET data for 
%V1494~Aql 1999\#2, to K. Page for sending us the data for V745~Sco (2014),
%and to the late A. Cassatella for providing us with  
%UV 1455 \AA~data for {\it IUE} novae.
     We thank
the American Association of Variable Star Observers
(AAVSO) and the Variable Star Observers League of Japan (VSOLJ)
for the archival data of V339~Del.
%PW~Vul, V5558~Sgr, HR~Del, and V723~Cas.
%%This research has been supported in part by Grants-in-Aid for
%%Scientific Research (15K05026, 16K05289)
%%from the Japan Society for the Promotion of Science.
We also thank the ex-students who participated in the intensive
observation at OKU, especially, Minami Matsuura, Miho Kawabata,
Naoto Kojiguchi, and Yuki Sugiura.
KM acknowledges JSPS KAKENHI grant number JP19K03930.
\end{acknowledgments}

\end{document}